\documentclass[aps,prl,reprint,preprintnumbers]{revtex4-1}
\usepackage{url}
\usepackage{textcase}
\usepackage{bm}
\usepackage{amsmath}
\usepackage{amsfonts}
\usepackage{amssymb}
\usepackage{hyperref}
\usepackage[table]{xcolor}
\usepackage{multirow}
\usepackage{bbm}
\usepackage{graphicx}
\usepackage{tensor}
\usepackage[caption=false]{subfig}
\usepackage{array}
\usepackage[normalem]{ulem}
\usepackage{float}
\usepackage{fancyhdr}

\usepackage{pdfpages}
\makeatletter
\AtBeginDocument{\let\LS@rot\@undefined}
\makeatother

\usepackage{natbib}
\renewcommand{\Re}{\mathrm{Re}}
\renewcommand{\vec}[1]{\boldsymbol{#1}}

\begin{document}

\title{Scale anomaly of a Lifshitz scalar: a universal quantum phase transition to discrete scale invariance}

\date{\today}
\author{Daniel K.~Brattan}
\email{danny.brattan@gmail.com}
\affiliation{Interdisciplinary Center for Theoretical Study, University of Science and Technology of China, 96 Jinzhai Road, Hefei, Anhui, 230026 PRC.}
\author{Omrie Ovdat}
\email{somrie@campus.technion.ac.il}
\affiliation{Department of Physics, Technion Israel Institute of Technology, Haifa 3200003, Israel.}
\author{Eric Akkermans}
\email{eric@physics.technion.ac.il}
\affiliation{Department of Physics, Technion Israel Institute of Technology, Haifa 3200003, Israel.}

\begin{abstract}
{\ We demonstrate the existence of a universal transition from a continuous scale invariant phase to a discrete scale invariant phase for a class of one-dimensional quantum systems with anisotropic scaling symmetry between space and time. These systems describe a Lifshitz scalar interacting with a background potential. The transition occurs at a critical coupling $\lambda_{c}$ corresponding to a strongly attractive potential.}
\end{abstract}

\preprint{USTC-ICTS-17-08} 

\maketitle

{\ Some of the most intriguing phenomena resulting from quantum physics are the violation of classical symmetries, collectively referred to as anomalies \cite{PhysRev.177.2426,Bell1969,PhysRevD.34.674,1993AmJPh..61..142H}. One class of anomalies describes the breaking of continuous scale symmetry at the quantum level. A remarkable example of these ``scale anomalies'' occurs in the case of a non-relativistic particle in the presence of an attractive, inverse square potential \cite{Case1950,deAlfaro1976,landau1991quantum,Camblong:2000ec,PhysRevD.68.025006,Hammer:2005sa,Braaten2004,Kaplan:2009kr}, which describes the Efimov effect~\cite{efimov1970energy,Efimov1971,Braaten2006259} and plays a role in various other systems~\cite{levy1967electron,PhysRevB.46.12664,PhysRevD.48.5940,CamblongEpeleFanchiottiEtAl2001,nisoli2014attractive,Govindarajan:2000ag,Camblong:2003mz,Bellucci200399}. Classically scale invariant \cite{jackiw1995diverse}, any system described by the Hamiltonian, $ \hat{H}_S = p^2/2m-\lambda/r^2 $, exhibits an abrupt transition in the spectrum at $2 m \lambda_c = (d-2)^2/4$ \footnote{Thus, for $d=2$, an attractive potential is always overcritical.} where $d$ is the space dimension. For $\lambda < \lambda_c$, the spectrum contains no bound states close to $E=0$, however, as $\lambda$ goes above $\lambda_c$, an infinite series of bound states appears. Moreover, in this ``over-critical'' regime, the states arrange themselves in an unanticipated geometric series $E_n \propto \exp{(-2\pi n/\sqrt{\lambda-\lambda_{c}})}$, $n \in \mathbb{Z}$, accumulating at $E = 0$. The existence and geometric structure of such levels do not rely on the details of the potential close to its source and is a signature of residual discrete scale invariance since $\{E_n\} \rightarrow \{\exp{(-2\pi/\sqrt{\lambda-\lambda_{c}}))} E_n\} = \{E_{n}\}$. Thus, $H_S$ exhibits a quantum phase transition at $\lambda_c$ between a continuous scale invariant (CSI) phase and a discrete scale invariant phase (DSI). This transition has been associated with Berezinskii-Kosterlitz-Thouless (BKT) transitions \cite{PhysRevB.46.12664,Kaplan:2009kr,Jensen:2010ga,Jensen:2010vx,Jensen:2011af,2014JSP...156..268D,Gies:2015hia}.}

{\ Another system - the charged and massless Dirac fermion in an attractive Coulomb potential $\hat{H}_D = \gamma^0 \gamma^j p_j - \lambda/r$ \cite{OvdatMaoJiangEtAl2017} - also belongs to the same universal class of systems with these abrupt transitions. The similarity between the spectra and transition of these Dirac and Schr\"{o}dinger Hamiltonians motivates the study of whether a transition of this sort is possible for a generic scale invariant system. Specifically, these different Hamiltonians share a similar property -  the power law form of the corresponding potential matches the order of the kinetic term. In this paper, we examine whether this property is a sufficient ingredient by considering a generalised class of one dimensional Hamiltonians
  \begin{eqnarray}
    \label{Eq:Hamiltonian intro}
    \hat{H}_{N} = (p^{2})^{N} - \frac{\lambda_{N}}{x^{2N}} \; , 
  \end{eqnarray}
where $N$ is an integer and $\lambda_{N}$ a real coupling, and study if they exhibit a transition of the same universality class as $\hat{H}_{S}=\hat{H}_{1}, \hat{H}_{D}$.}

{\ Hamiltonian \eqref{Eq:Hamiltonian intro} describes a system with non-quadratic anisotropic scaling between space and time for $N>1$. This ``Lifshitz scaling symmetry'' \cite{Alexandre:2011kr}, manifest in \eqref{Eq:Hamiltonian intro}, can be seen for example at the finite temperature multicritical points of certain materials \cite{PhysRevLett.35.1678,PhysRevB.23.4615} or in strongly correlated electron systems \cite{PhysRevB.69.224415,PhysRevB.69.224416,Ardonne:2003wa}. It may also have applications in particle physics \cite{Alexandre:2011kr}, cosmology \cite{Mukohyama:2010xz} and quantum gravity \cite{PhysRevD.57.971,Kachru:2008yh,Horava:2009if,Horava:2009uw,Gies:2016con}. The non-interacting mode ($\lambda_{N}=0$) can also appear very generically, for example in non-relativistic systems with spontaneous symmetry breaking \cite{Brauner:2010wm}.}

{\ Generalising \eqref{Eq:Hamiltonian intro} to higher dimensional flat or curved spacetimes also introduces intermediate scale invariant terms which are products of radial derivatives and powers of the inverse radius. For $d>1$ the potential in \eqref{Eq:Hamiltonian intro} can be generated by considering a Lifshitz scale invariant system with charge and turning on a background gauge field \cite{Horava:2008jf,Das:2009fb,Alexandre:2011kr,Farias:2011aa} consisting of the appropriate multipole moment. The procedures we use throughout this paper are readily extended to these situations and the simple model \eqref{Eq:Hamiltonian intro} is sufficient to capture the desired features.}

\newcolumntype{C}{>{\centering\let\newline\\\arraybackslash\hspace{0pt}}m{2.6cm}}
\begin{table*}[!t]
  \centering
  \begin{ruledtabular}
  \begin{tabular}{rCCCCCC}
    \; $N$ \; 		      &  \multicolumn{2}{c}{$1$} 	    & \multicolumn{2}{c}{$2$} 				              
			      &  \multicolumn{2}{c}{$3$}                                 \\ \hline
    \; $\lambda$ \;           & $-\frac{3}{4} < \lambda_{1} $ & $ \lambda_{1} < -\frac{3}{4} $ & $\frac{105}{16} < \lambda_{2} $ & $ -45 < \lambda_{2} < \frac{105}{16}  $ 
			      & $ 693 \lesssim \lambda_{3} $  &  $ -162 \lesssim \lambda_{3}  \lesssim  693$ \\ \hline
    \; extension    \; & \cellcolor{lightgray} $U(1)$ 	    & \; self-adjoint \;	             	& \cellcolor{lightgray} $U(1)$                   	
			      & $U(2)$      & \cellcolor{lightgray} $U(1)$ 	    & $U(3)$
  \end{tabular}
  \end{ruledtabular}
  \caption{The regimes of self-adjoint extension parameter with $N=1,2,3$ for some values of $\lambda_{N}$. The bounds for $N=1,2$ are exact while those for $N=3$ are approximate and determined by numerically solving the indicial equation \eqref{Eq:indicialeqn}. A table showing $\lambda_{N}$, for $N=1,2,3$, to larger negative values can be found in the supplementary material.}
  \label{tab:SAEparamregimes}
  \vspace{-1em}
\end{table*}

{\ Our main results are summarized as follows. In accordance with the $N = 1$ case, there is a quantum phase transition at $\lambda_{N,c} \equiv (2N-1)!!^{2}/2^{2N}$ for all $N>1$ from a CSI phase to a DSI phase in the low energy regime $|E|^{1/2N}x_0 \ll 1$, where $x_0 > 0$ is a short distance cutoff. The CSI phase contains no bound states and the DSI phase is characterized by an infinite set of bound states forming a geometric series as given by equation \eqref{Eq:selfadjointenergies}. The transition and $\lambda_{N,c}$ value are independent of the short distance physics characterised by the boundary condition at $x = x_0$. For $\left( \lambda_N - \lambda_{N,c} \right) \rightarrow 0^+$, the analytic behaviour of the spectrum is characteristic of the BKT scaling in analogy with $N = 1$ \cite{PhysRevB.46.12664,Kaplan:2009kr} and as shown by equation \eqref{Eq:BKT scaling}. We analyse the $x_0 = 0$ case, obtain its self adjoint extensions and spectrum and obtain similar results.}

\section{The model}

{\ Corresponding to \eqref{Eq:Hamiltonian intro} is the action of a complex scalar field in $(1+1)$-dimensions:
  \begin{eqnarray}
	\label{Eq:Action}
    &\;& \int dt \int_{x=x_{0}}^{\infty} dx \; \frac{i}{2} \left( \Psi^{*} \partial_{t} \Psi - \mathrm{c.c.} \right) 
	 - \left| \partial_{x}^{N} \Psi \right|^{2} + \frac{\lambda_{N}}{x^{2N}} \left| \Psi \right|^2 \; , \nonumber
  \end{eqnarray}
where $\mathrm{c.c.}$ indicates the complex conjugate. This field theory has manifest Lifshitz scaling symmetry, $(t,x) \mapsto (\Lambda^{2N} t, \Lambda x)$ when $x_{0} \rightarrow 0$. The scaling exponent of $\Lambda^{2}$ is called the ``dynamical exponent'' and has value $N$ in this case. The action represents a Lifshitz scalar with a single time derivative and can be recovered as the low energy with respect to mass limit of a charged, massive Lifshitz scalar which is quadratic in time derivatives \cite{Pal:2016rpz}. The eigenstates of Hamiltonian \eqref{Eq:Hamiltonian intro} are given by stationary solutions of the subsequent equations of motion \footnote{Additional boundary terms will be required to make the variation of the action vanish on boundary conditions that make the corresponding Hamiltonian operator self-adjoint \cite{Asorey:2006pr}.}.}
 
{\ Consider the case $x_0 < x < \infty$ with $x_0 = 0$. The classical scaling symmetry of \eqref{Eq:Hamiltonian intro} implies that if there is one negative energy bound state then there is an unbounded continuum. Thus, the Hamiltonian is non-self-adjoint \cite{Bonneau:1999zq,2015AnHP...16.2367I}. The origin of this phenomenon, already known from the $N=1$ case, is the strong singularity of the potential at $x=0$. To remedy this problem, the operator can be made self-adjoint by applying boundary conditions on the elements of the Hilbert space through the procedure of self-adjoint extension \cite{Gitman:2009era}. Alternatively, a suitable cutoff regularisation at $x_0>0$ can be chosen to ensure self-adjointness as well as bound the spectrum from below by an intrinsic scale \cite{Camblong:2000ec} leaving some approximate DSI at low energies.}
 
{\ For $N=1$ in \eqref{Eq:Hamiltonian intro}, the continuous scaling invariance of $\hat{H}_{1}$ is broken anomalously as a result of restoring self-adjointness. In particular, for $\lambda_{1}>1/4$, one obtains an energy spectrum given by $E_{1} = - E_{0} \exp \left( - (2 \pi n) / \sqrt{\lambda_{1}-\lambda_{c}} \right)$ for $n \in \mathbbm{Z}$. The energies are related by a discrete family of scalings which manifest a leftover DSI in the system.}

{\ In what follows, we shall demonstrate that in the case of large positive $\lambda_{N}$ (attractive potential) one finds, for all $N$, a regime with a geometric tower of states using the method of self-adjoint extension. Subsequently, we bound the resulting spectrum from below, while maintaining DSI at low energies, by keeping $x_{0} > 0$ and applying generic (self-adjoint) boundary conditions at a cut-off point. We obtain that a transition from CSI to DSI is a generic feature in the landscape of these Hamiltonians \eqref{Eq:Hamiltonian intro}, independent of the choice of boundary condition and in a complete analogy with the $N=1$ case \cite{Case1950,deAlfaro1976,landau1991quantum,Camblong:2000ec,PhysRevD.68.025006,Hammer:2005sa,Braaten2004,Kaplan:2009kr}. By analytically solving the eigenvalue problem for all $N$, we obtain an expression for the critical $\lambda_{N}$ and for the resulting DSI spectrum in the over-critical regime.}

\section{Self-adjoint extension}

{\ To determine whether the Hamiltonians \eqref{Eq:Hamiltonian intro} can be made self-adjoint one can apply von Neumann's procedure \cite{Meetz1964,Bonneau:1999zq,Gitman:2009era,gitman2012self,2015AnHP...16.2367I}. This consists of counting the normalisable solutions to the energy eigenvalue equation with unit imaginary energies of both signs. When there are $M$ linearly independent solutions of positive and negative sign respectively, there is a $U(M)$ parameter family of conditions at $x = 0$ that can make the operator self-adjoint. If $M=0$, it is essentially self-adjoint and if the number of positive and negative solutions do not match then the operator cannot be made self-adjoint.}

\begin{figure}[!t]
  \centering
  \includegraphics[width=\columnwidth]{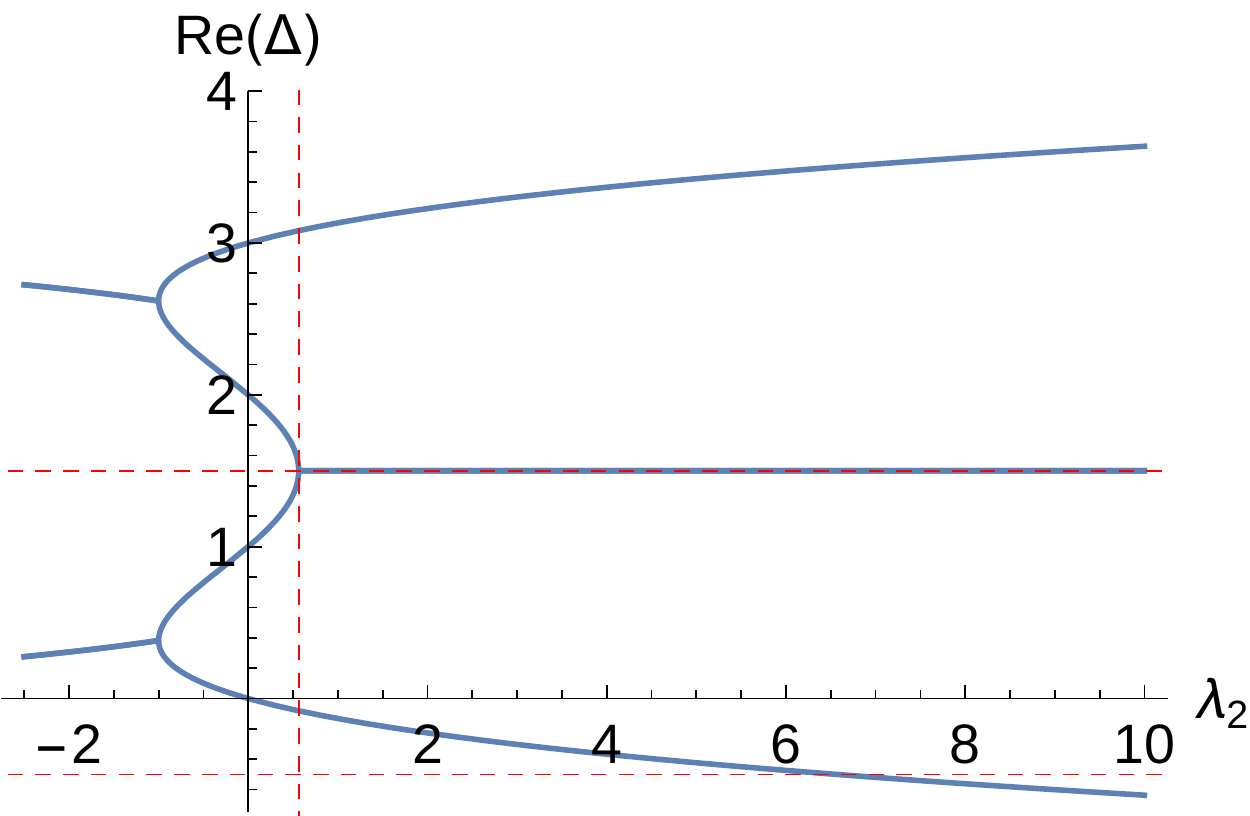}
  \caption{A plot of the real part of the solution to the indicial equation \eqref{Eq:indicialeqn} for $N=2$ against $\lambda_{2}$. The dotted, red vertical line indicates where $\lambda_{2}=9/16$ above which the first pair of complex roots appears. The lowermost horizontal red line indicates where one of the roots becomes non-normalisable at positive $\lambda_{2}$ (i.e.~$\lambda_{2}=105/16$) while the uppermost indicates the symmetry line responsible for pairing roots i.e.~$\Delta=3/2$.}
  \label{fig:N=4realpartofroots}
  \vspace{-1em}
\end{figure}

{\ For every $N$, there are $N$ exponentially decaying solutions at infinity for $E=\pm i$. Normalisability then depends on their $x \rightarrow 0$ behaviour which requires a complete solution of the eigenfunctions. The energy eigenvalue equation from \eqref{Eq:Hamiltonian intro} has an analytic solution in terms of generalised hypergeometric functions \cite{luke1969special} (see the supplementary material) for arbitrary complex energy and $\lambda_{N}$. Near $x = 0$, the analytic solutions are characterised by the roots of the indicial equation,
  \begin{eqnarray}
    \label{Eq:indicialeqn}
    \lambda_{N} = (-1)^{N} \Delta( \Delta - 1) (\Delta-2) \ldots (\Delta - 2 N + 1 ) \; , \qquad
  \end{eqnarray}
which can be obtained by inserting $\Psi(x) = x^{\Delta} (1 + \mathcal{O}(x^{2N}) )$ into the eigenvalue equation and solving for $\Delta$. We label the $2N$ solutions of \eqref{Eq:indicialeqn} as $\Delta_{i}, i =1, \ldots, 2N$ and order by ascending real part followed by ascending imaginary part. Since \eqref{Eq:indicialeqn} is real and symmetric about $N-1/2$, all roots can be collected into pairs which sum to $2N-1$ (see fig.~\ref{fig:N=4realpartofroots}). In addition, because $N-1/2 \geq -1/2$, there can be at most $N$ roots with $\Re(\Delta_{i}) \leq -1/2$ . Thus, near $x = 0$, a generic combination of the $N$ decaying solutions has the form: 
  \begin{eqnarray}
    \label{Eq:GenericNearOriginExpansion}
    \Psi(x) &=& \sum_{i=1}^{N} \left[ \left( \epsilon x \right)^{\Delta_{i}} \phi_{i} + \ldots + \left( \epsilon x \right)^{\Delta_{i+N}}  O_{i} + \ldots \right]
    \; , \qquad
  \end{eqnarray}
where $E$ is the energy, $\epsilon \equiv |E|^{1/2N}$ and $\phi_{i},O_{i}$ are complex numbers \footnote{We have also assumed that the $\Delta_{i}$ do not differ by integer powers to avoid the complication of logarithmic terms in the Frobenius series represented by \eqref{Eq:GenericNearOriginExpansion}.}. We have distinguished modes with the potential to be divergent as $x\rightarrow 0$ by the coefficients $\phi_{i}$. Taking ratios of $O_{i}$ and $\phi_{i}$ yields $N$ independent dimensionless scales.}

{\ To obtain the number of independent normalisable solutions $M$, we need to count the number $m$ of roots $\Delta_{i}$ with real parts less than $-1/2$ as $\lambda_{N}$ varies. When any root violates this bound we must take a linear combination of our $N$ decaying functions to remove it from the near origin expansion and as a result $M = N - m$. This is easily accomplished by examining the analytic solutions in an expansion about $x=0$.}

{\ Generically, for $\lambda_{N}$ small in magnitude, $m =0$, and there is an $U(N)$ self-adjoint extension \footnote{In fact, for $\lambda_{N}$ small enough the $\Delta_{i}$ are real and one can find isolated bound states for some choices of the parameter $U(N)$.}. As $\lambda_{N}$ increases in absolute value and the real part of some roots go below $-1/2$ the dimension of the self-adjoint parameter decreases. Eventually $\hat{H}_{N}$ either becomes essentially self-adjoint (negative $\lambda_{N}$, repulsive potential) or has a $U(1)$ self-adjoint extension (positive $\lambda_{N}$, attractive potential). The different regimes are summarised for $N=1,2,3$ in table \ref{tab:SAEparamregimes}.}

{\ Now that we have determined when our operator can be made self-adjoint we would like to obtain its energy spectrum. To that purpose, we define the boundary form, $\left[ \Psi, \Phi \right](x)$, by computing the difference $\hat{H}_N - \hat{H}^\dagger _N$ :
  \begin{eqnarray}
   \label{Eq:differenceHamiltonians}
   \langle \Phi | \hat{H}_{N} | \Psi \rangle &-& \langle \Phi | \hat{H}^{\dagger}_{N} | \Psi \rangle = \left[ \Phi, \Psi \right](\infty) - \left[ \Phi, \Psi \right](x_{0}) \; , \qquad \\
   \langle \Phi | \hat{H}_{N} | \Psi \rangle &=& \int_{x=x_{0}}^{\infty} dx \; \Phi^{*}(x) \hat{H}_{N} \Psi(x) \; , 
  \end{eqnarray}
where
  \begin{eqnarray}
   \label{Eq:asymmetryform}
   \left[ \Phi, \Psi \right](x) = \sum_{j=1}^{2N} (-1)^{N+j-1} d_{x}^{j-1} \Phi^{*}(x) d_{x}^{2N-j} \Psi(x) \;  \qquad
  \end{eqnarray}
and the boundary form is calculated by moving the derivatives in $\langle \Phi | \hat{H}_{N} | \Psi \rangle$ to obtain $\langle \Phi | \hat{H}_{N}^{\dagger} | \Psi \rangle$ (see supplementary material). Setting $\Phi=\Psi$ and using the time dependent Schr\"{o}dinger equation it is straightforward to show that 
  \begin{eqnarray}
    \partial_{t} \rho = \partial_{x} \left[ \Psi, \Psi \right] \; , \qquad \rho = |\Psi|^2 \; , \qquad
  \end{eqnarray}
so that the boundary form can be interpreted as the value of the probability current at $x=x_{0}$.}

{\ We are interested in evaluating \eqref{Eq:differenceHamiltonians} on the energy eigenfunctions and require it to vanish when taking $x_{0} \rightarrow 0$ as this fixes any remaining free parameters of a general solution. For $E<0$ there are $N$ exponentially decaying solutions at infinity. A general energy eigenfunction is a sum of these and thus the boundary form \eqref{Eq:differenceHamiltonians} evaluated at infinity is zero. For $\hat{H}_{N}$ to be self-adjoint it is necessary to impose the same boundary conditions on the wavefunctions and their adjoints while ensuring that \eqref{Eq:asymmetryform} vanishes.}

\begin{figure}[!t]
  \centering
  \includegraphics[width=\columnwidth]{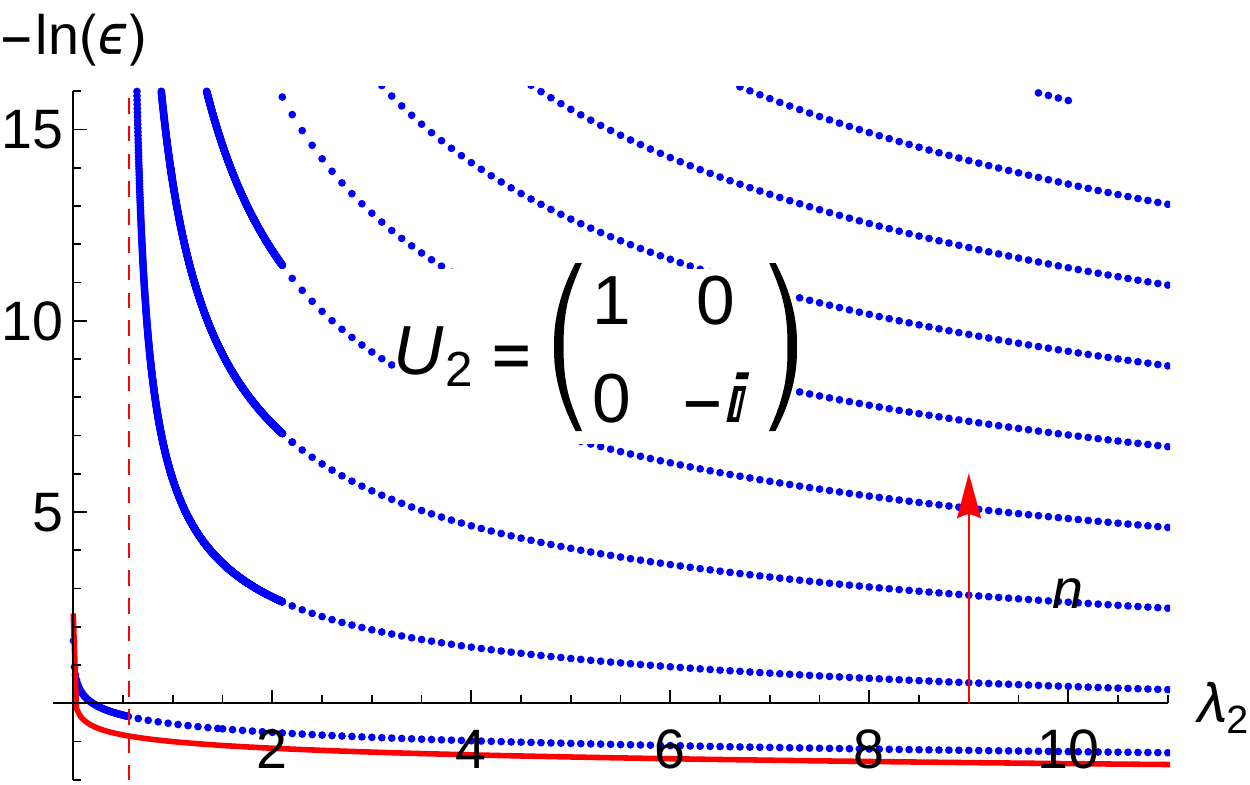}
  \caption{A plot of the logarithm of $\epsilon$ against $\lambda_{2}$ for a cut-off position $x_{0} = e^{-1}$ and boundary condition displayed. For $\lambda_{2}<\lambda_{2,c}=9/16$ there are no bound states satisfying $\epsilon x_{0} \ll 1$ (although there is an isolated bound state outside this energy bound). The solid red line indicates the lower bound on the negative energies as discussed in the supplementary material. The dotted red line at $\lambda_{2,c}$ indicates where the first pair of complex roots appears, above which we can see the geometric tower abruptly appearing from $\epsilon=0$.}
  \label{fig:N=4cutoff}
  \vspace{-1em}
\end{figure}

{\ By examining \eqref{Eq:indicialeqn} it can be seen that for $\lambda_{N} > \lambda_{N,c} \equiv (2N-1)!!^{2}/2^{2N}$ there is a pair of complex roots whose real part is fixed to be $N-1/2$, i.e.,
  \begin{eqnarray}
   \label{Eq:SpecialRoots}
   \Delta_{N}&=&\Delta^{*}_{N+1}= (N-1/2) - i \nu_{N}(\lambda_{N}) \; , \\
   \label{Eq:nuintermsoflambdacrit}
   \nu_{N}(\lambda_{N}) &=& \sqrt{\lambda_{N} - \lambda_{N,c}} \left( \alpha_{N}^{-1} + \mathcal{O}(\lambda_{N} - \lambda_{N,c}) \right) \; , \qquad
  \end{eqnarray}
where $\alpha_{N}$ is a constant given explicitly in the supplementary material. At sufficiently positive $\lambda_{N}$, $\Delta_{1},\dots,\Delta_{N-1}$ are not normalisable and the pair \eqref{Eq:SpecialRoots} are the leading normalisable roots. This is illustrated in fig.~\ref{fig:N=4realpartofroots} for $N=2$. Removing the $N-1$ non-normalisable roots yields a single decaying wavefunction with arbitrary energy. As such, \eqref{Eq:GenericNearOriginExpansion} becomes 
  \begin{eqnarray}
   \label{Eq:OurWavefunction}
   \Psi(x) &=& \phi_{N} \left( \epsilon x \right)^{N-1/2-i\nu_{N}} + O_{1} \left( \epsilon x \right)^{N-1/2 + i \nu_{N}} \nonumber \\
	   &\;& + \sum_{j \geq 2}  O_{j} \left( \epsilon x \right)^{\Delta_{j+N}} + \ldots \; , \qquad
  \end{eqnarray}
where the $N$ dimensionless and energy independent scales, $O_{j}/\phi_{N}$, are now fixed. Substituting \eqref{Eq:OurWavefunction} for two energies $E$ and $\tilde{E}$ into \eqref{Eq:asymmetryform} yields a term $\propto ( |E/\tilde{E}|^{2 i \nu_{N}/N} - |O_{1}/\phi_{N}|^2 )$ as $x_{0} \rightarrow 0$. Thus, the self-adjoint boundary condition is equivalently an energy quantisation condition relating an arbitrary reference energy $E_{0}>0$ to a geometric tower of energies:
  \begin{eqnarray}
   \label{Eq:selfadjointenergies}
    E_{n} &=& - E_{0} e^{- \frac{N \pi n}{\nu_{N}}} \; , \; \; n \in \mathbbm{Z} \; . 
  \end{eqnarray}
The reference energy $E_{0}$ is a free parameter and can be chosen arbitrarily.}

\section{Cut-off regularisation}

{\ We shall now consider $x_{0}>0$ and choose boundary conditions to give a lower bound to the energy spectrum while preserving approximate DSI near zero energy. We will impose the most general boundary condition on the cut-off point consistent with unitary time evolution \footnote{The Hamiltonian is to be made self-adjoint in the portion of the half-line considered.}. As the $M=N$ decaying solutions are finite at all points $x_{0}>0$, independent of $\lambda_{N}$, this general boundary condition corresponds to an $U(N)$ self-adjoint extension.}

{\ To obtain the boundary condition, we diagonalize \eqref{Eq:asymmetryform} by defining $\Psi_{k}^{\pm}(x_{0})$,
  \begin{eqnarray}
   \label{Eq:diagonalise}
	 x_{0}^{k-1} d_{x}^{k-1} \Psi(x_{0})
     &=& \Psi_{k}^{+}(x_{0}) + \Psi_{k}^{-}(x_{0}) \; ,  \\
	 x_{0}^{2N-k} d_{x}^{2N-k} \Psi(x_{0})
     &=& e^{i \pi (k-\frac{1}{2})} \left[ \Psi_{k}^{+}(x_{0}) - \Psi_{k}^{-}(x_{0}) \right] \qquad 
  \end{eqnarray}
for $1 \leq k \leq N$. After this redefinition, \eqref{Eq:asymmetryform} evaluated at $x=x_{0}$ becomes proportional to
  \begin{eqnarray}
	\label{Eq:antisymmetryform}
	\vec{\Phi}^{+}(x_{0})^{\dagger} \cdot  \vec{\Psi}^{+}(x_{0}) - \vec{\Phi}^{-}(x_{0})^{\dagger} \cdot \vec{\Psi}^{-}(x_{0}) \; .  \qquad
  \end{eqnarray}
The vanishing of \eqref{Eq:antisymmetryform} can be achieved by setting
  \begin{eqnarray}
   \label{Eq:GenericUnBoundaryCondition}
   \vec{\Psi}^{+}(x_{0}) = U_{N} \vec{\Psi}^{-}(x_{0}) \; , 
  \end{eqnarray}
for some arbitrary unitary matrix: $U_{N}$ \cite{Gitman:2009era,9780817646622}. This additionally ensures that elements of the space of wavefunctions and its adjoint have the same boundary conditions, making $\hat{H}_{N}$ self-adjoint on said space. The matrix $U_{N}$ can only be specified by supplying additional physical information beyond the form of the Hamiltonian.}

\begin{figure}[!t]
  \centering
  \includegraphics[width=\columnwidth]{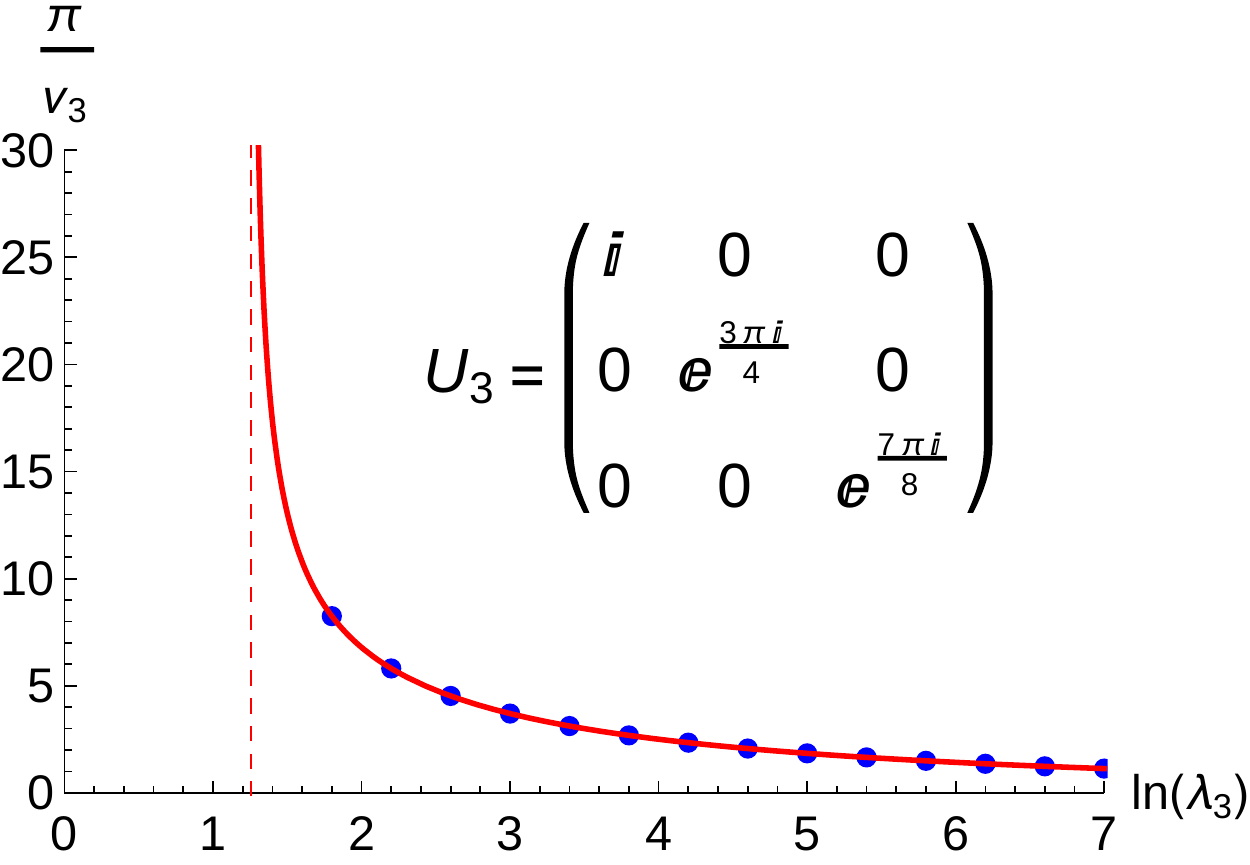}
  \caption{A plot of $\pi / \nu_{3}$ against $\ln \lambda_{3}$ for a cut-off position $x_{0} = e^{-1}$ and boundary condition displayed. The solid red line indicates the numerical result from solving \eqref{Eq:indicialeqn} for the roots defined in \eqref{Eq:SpecialRoots}. The blue dots are calculated by numerically determining the gradient of $\ln E_{n}/E_{n+1}$ against $n$ for several $n$ corresponding to $\epsilon x_{0} \ll 1$. The red dotted line indicates the critical $\lambda_{3}$.}
  \label{fig:N=6cutoff}
  \vspace{-1em}
\end{figure}

{\ As an illustration of the appearance of the geometric tower at $N>1$, consider figs.~\ref{fig:N=4cutoff} and \ref{fig:N=6cutoff}. The former plots $\epsilon$ for $N=2$ against $\lambda_{2}$. It is plain that as soon as $\lambda_{2}>9/16$ (the dotted red line) there is a sudden transition from no bound states satisfying $\epsilon \ll x_0^{-1}$  (and one isolated bound state), to a tower of states. Similarly fig.~\ref{fig:N=6cutoff} plots the logarithm of $E_{n}/E_{n+1}$ for $N=3$ as a function of $\lambda_{3}$ at low $\epsilon x_{0}$. The result, shown by the blue points in fig.~\ref{fig:N=6cutoff}, is a good match with $\pi/\nu_{3}$ with $\nu_{3}$ defined by \eqref{Eq:SpecialRoots}.}

{\ For general $N$, $\lambda_{N} > \lambda_{N,c}$ and small enough energies we shall now argue that one always finds DSI with the scaling defined in \eqref{Eq:selfadjointenergies} using a small $\epsilon$ expansion. Determining the energy eigenstates analytically for arbitrary boundary conditions is made difficult for $N>1$ due to the presence of multiple distinct complex roots in the small energy expansion. However, we shall show that one pair makes a contribution that decays more slowly as $\epsilon \rightarrow 0$ than any other and derive an approximation in this limit.}

{\ Consider \eqref{Eq:GenericNearOriginExpansion} evaluated at $x \sim x_{0}$ which is a good approximation to the wavefunction when $\epsilon x_{0} \ll 1$. Imposing decay fixes $N$ of the $2N$ free parameters. Without loss of generality we can take them to be $O_{i}$ so that $O_{i}=G\indices{_{i}^{j}} \phi_{j}$ for some complex $(N\times N)$-matrix $G$. The $N$ remaining free parameters $\phi_{i}$ will be fixed by wavefunction normalisation and boundary conditions at the cut-off point \eqref{Eq:GenericUnBoundaryCondition}. The generic dependence of $\phi_i$ on $\epsilon, x_0$ for $\epsilon x_0 \ll 1$ is extractable. Applying the redefinition $\tilde{\phi}_{i} \equiv \phi_{i} (\epsilon x_0)^{\Delta_{i}}$ implies 
  \begin{eqnarray}
   O_{i} (\epsilon x_0)^{\Delta_{i+N}} \rightarrow G\indices{_{i}^{j}} (\epsilon x_0)^{\Delta_{i+N}-\Delta_{j}}  \tilde{\phi}_{j} \; . \qquad
  \end{eqnarray} 
Given our canonical ordering of the roots and that $\lambda>\lambda_{c}$ we have $\Re(\Delta_{i+N}-\Delta_{j}) \geq 0$ with equality only when $i=1$ and $j=N$. For $\epsilon x_0 \ll 1$ the leading contributions to the wavefunction \eqref{Eq:GenericNearOriginExpansion} have the form 
  \begin{eqnarray}
   \label{Eq:SmallEnergyExpansion}
    \Psi(x) &=& \sum_{i=1}^{N} \tilde{\phi}_{i} \left(\frac{x}{x_0}\right)^{\Delta_{i}}
		+ G\indices{_{1}^{N}} \tilde{\phi}_{N} \left( \epsilon x_{0} \right)^{2i\nu_{N}} \left(\frac{x}{x_{0}}\right)^{\Delta_{N+1}} \nonumber
  \end{eqnarray}
where $\epsilon x_0$ only enters the leading term through a phase and all other contributions to $O_{i}$ from the $\tilde{\phi}_{i}$ drop out as they come with $\epsilon x_0$ to a real positive power. The displayed terms above are the relevant ones at low energies for solving \eqref{Eq:GenericUnBoundaryCondition}. Moreover these leading terms are invariant under the discrete scaling transformation and thus we have DSI. As a result, applying \eqref{Eq:GenericUnBoundaryCondition} will necessarily give the energy spectrum \eqref{Eq:selfadjointenergies} for $\epsilon x_{0} \ll 1$ with $E_{0}$ a number depending on $U(N)$ and the cut-off $x_{0}$.}

{\ We can use our expression \eqref{Eq:nuintermsoflambdacrit} for $\nu_{N}$ in terms of $\lambda_{N}$ to find:
  \begin{eqnarray}
%     E_{n} &=& -E_{0} e^{- \frac{N \pi n \alpha_{N}}{\sqrt{\lambda_{N}-\lambda_{N,c}}}} \left( 1 + \mathcal{O}(\lambda_{N}-\lambda_{N,c}) \right) \; , \qquad
    \frac{E_{n}}{E_{0}} &=& - \exp \left( - \frac{N \pi n \alpha_{N}}{ \sqrt{\lambda_{N} - \lambda_{N,c}} \left( 1 + \mathcal{O}(\lambda_{N}-\lambda_{N,c}) \right)  } \right) \; , \qquad
    \label{Eq:BKT scaling}
  \end{eqnarray}
characteristic of the BKT scaling for $\lambda_{N} \rightarrow \lambda_{N,c}$.}

{\ With the above considerations we can say that a CSI to DSI transition, at $\epsilon x_{0} \ll 1$, is a generic feature of our models is independent of the completion of the potential near the origin. This is in complete analogy with the $N=1$ case. Thus, the Hamiltonian \eqref{Eq:Hamiltonian intro} need only be effective for the consequences of DSI to be relevant.}

\begin{acknowledgments}
{\ The work of DB was supported in part by the Israel Science Foundation under grant 504/13 and is currently supported by a key grant from the NSF of China with Grant No: 11235010. This work was also supported by the Israel Science Foundation Grant No.~924/09. DB would like to thank the Technion and University of Haifa at Oranim for their support. DB would also like to thank Matteo Baggioli and Yicen Mou for reading early drafts.}
\end{acknowledgments}

%%%%%%%%%%%% Bibliography %%%%%%%%%%%%

\bibliography{references}

%merlin.mbs apsrev4-1.bst 2010-07-25 4.21a (PWD, AO, DPC) hacked
%Control: key (0)
%Control: author (8) initials jnrlst
%Control: editor formatted (1) identically to author
%Control: production of article title (-1) disabled
%Control: page (0) single
%Control: year (1) truncated
%Control: production of eprint (0) enabled
\begin{thebibliography}{60}%
\makeatletter
\providecommand \@ifxundefined [1]{%
 \@ifx{#1\undefined}
}%
\providecommand \@ifnum [1]{%
 \ifnum #1\expandafter \@firstoftwo
 \else \expandafter \@secondoftwo
 \fi
}%
\providecommand \@ifx [1]{%
 \ifx #1\expandafter \@firstoftwo
 \else \expandafter \@secondoftwo
 \fi
}%
\providecommand \natexlab [1]{#1}%
\providecommand \enquote  [1]{``#1''}%
\providecommand \bibnamefont  [1]{#1}%
\providecommand \bibfnamefont [1]{#1}%
\providecommand \citenamefont [1]{#1}%
\providecommand \href@noop [0]{\@secondoftwo}%
\providecommand \href [0]{\begingroup \@sanitize@url \@href}%
\providecommand \@href[1]{\@@startlink{#1}\@@href}%
\providecommand \@@href[1]{\endgroup#1\@@endlink}%
\providecommand \@sanitize@url [0]{\catcode `\\12\catcode `\$12\catcode
  `\&12\catcode `\#12\catcode `\^12\catcode `\_12\catcode `\%12\relax}%
\providecommand \@@startlink[1]{}%
\providecommand \@@endlink[0]{}%
\providecommand \url  [0]{\begingroup\@sanitize@url \@url }%
\providecommand \@url [1]{\endgroup\@href {#1}{\urlprefix }}%
\providecommand \urlprefix  [0]{URL }%
\providecommand \Eprint [0]{\href }%
\providecommand \doibase [0]{http://dx.doi.org/}%
\providecommand \selectlanguage [0]{\@gobble}%
\providecommand \bibinfo  [0]{\@secondoftwo}%
\providecommand \bibfield  [0]{\@secondoftwo}%
\providecommand \translation [1]{[#1]}%
\providecommand \BibitemOpen [0]{}%
\providecommand \bibitemStop [0]{}%
\providecommand \bibitemNoStop [0]{.\EOS\space}%
\providecommand \EOS [0]{\spacefactor3000\relax}%
\providecommand \BibitemShut  [1]{\csname bibitem#1\endcsname}%
\let\auto@bib@innerbib\@empty
%</preamble>
\bibitem [{\citenamefont {Adler}(1969)}]{PhysRev.177.2426}%
  \BibitemOpen
  \bibfield  {author} {\bibinfo {author} {\bibfnamefont {S.~L.}\ \bibnamefont
  {Adler}},\ }\href {\doibase 10.1103/PhysRev.177.2426} {\bibfield  {journal}
  {\bibinfo  {journal} {Phys. Rev.}\ }\textbf {\bibinfo {volume} {177}},\
  \bibinfo {pages} {2426} (\bibinfo {year} {1969})}\BibitemShut {NoStop}%
\bibitem [{\citenamefont {Bell}\ and\ \citenamefont {Jackiw}(1969)}]{Bell1969}%
  \BibitemOpen
  \bibfield  {author} {\bibinfo {author} {\bibfnamefont {J.~S.}\ \bibnamefont
  {Bell}}\ and\ \bibinfo {author} {\bibfnamefont {R.}~\bibnamefont {Jackiw}},\
  }\href {\doibase 10.1007/BF02823296} {\bibfield  {journal} {\bibinfo
  {journal} {Il Nuovo Cimento A (1965-1970)}\ }\textbf {\bibinfo {volume}
  {60}},\ \bibinfo {pages} {47} (\bibinfo {year} {1969})}\BibitemShut {NoStop}%
\bibitem [{\citenamefont {Esteve}(1986)}]{PhysRevD.34.674}%
  \BibitemOpen
  \bibfield  {author} {\bibinfo {author} {\bibfnamefont {J.~G.}\ \bibnamefont
  {Esteve}},\ }\href {\doibase 10.1103/PhysRevD.34.674} {\bibfield  {journal}
  {\bibinfo  {journal} {Phys. Rev. D}\ }\textbf {\bibinfo {volume} {34}},\
  \bibinfo {pages} {674} (\bibinfo {year} {1986})}\BibitemShut {NoStop}%
\bibitem [{\citenamefont {{Holstein}}(1993)}]{1993AmJPh..61..142H}%
  \BibitemOpen
  \bibfield  {author} {\bibinfo {author} {\bibfnamefont {B.~R.}\ \bibnamefont
  {{Holstein}}},\ }\href {\doibase 10.1119/1.17328} {\bibfield  {journal}
  {\bibinfo  {journal} {American Journal of Physics}\ }\textbf {\bibinfo
  {volume} {61}},\ \bibinfo {pages} {142} (\bibinfo {year} {1993})}\BibitemShut
  {NoStop}%
\bibitem [{\citenamefont {Case}(1950)}]{Case1950}%
  \BibitemOpen
  \bibfield  {author} {\bibinfo {author} {\bibfnamefont {K.~M.}\ \bibnamefont
  {Case}},\ }\href {\doibase 10.1103/PhysRev.80.797} {\bibfield  {journal}
  {\bibinfo  {journal} {Phys. Rev.}\ }\textbf {\bibinfo {volume} {80}},\
  \bibinfo {pages} {797} (\bibinfo {year} {1950})}\BibitemShut {NoStop}%
\bibitem [{\citenamefont {de~Alfaro}\ \emph {et~al.}(1976)\citenamefont
  {de~Alfaro}, \citenamefont {Fubini},\ and\ \citenamefont
  {Furlan}}]{deAlfaro1976}%
  \BibitemOpen
  \bibfield  {author} {\bibinfo {author} {\bibfnamefont {V.}~\bibnamefont
  {de~Alfaro}}, \bibinfo {author} {\bibfnamefont {S.}~\bibnamefont {Fubini}}, \
  and\ \bibinfo {author} {\bibfnamefont {G.}~\bibnamefont {Furlan}},\ }\href
  {\doibase 10.1007/BF02785666} {\bibfield  {journal} {\bibinfo  {journal} {Il
  Nuovo Cimento A (1965-1970)}\ }\textbf {\bibinfo {volume} {34}},\ \bibinfo
  {pages} {569} (\bibinfo {year} {1976})}\BibitemShut {NoStop}%
\bibitem [{\citenamefont {Landau}(1991)}]{landau1991quantum}%
  \BibitemOpen
  \bibfield  {author} {\bibinfo {author} {\bibfnamefont {L.~D.}\ \bibnamefont
  {Landau}},\ }\href@noop {} {\emph {\bibinfo {title} {Quantum mechanics :
  non-relativistic theory}}}\ (\bibinfo  {publisher} {Butterworth-Heinemann},\
  \bibinfo {address} {Oxford Boston},\ \bibinfo {year} {1991})\BibitemShut
  {NoStop}%
\bibitem [{\citenamefont {Camblong}\ \emph {et~al.}(2000)\citenamefont
  {Camblong}, \citenamefont {Epele}, \citenamefont {Fanchiotti},\ and\
  \citenamefont {Garcia~Canal}}]{Camblong:2000ec}%
  \BibitemOpen
  \bibfield  {author} {\bibinfo {author} {\bibfnamefont {H.~E.}\ \bibnamefont
  {Camblong}}, \bibinfo {author} {\bibfnamefont {L.~N.}\ \bibnamefont {Epele}},
  \bibinfo {author} {\bibfnamefont {H.}~\bibnamefont {Fanchiotti}}, \ and\
  \bibinfo {author} {\bibfnamefont {C.~A.}\ \bibnamefont {Garcia~Canal}},\
  }\href {\doibase 10.1103/PhysRevLett.85.1590} {\bibfield  {journal} {\bibinfo
   {journal} {Phys. Rev. Lett.}\ }\textbf {\bibinfo {volume} {85}},\ \bibinfo
  {pages} {1590} (\bibinfo {year} {2000})},\ \Eprint
  {http://arxiv.org/abs/hep-th/0003014} {arXiv:hep-th/0003014 [hep-th]}
  \BibitemShut {NoStop}%
%%CITATION = HEP-TH/0003014;%%
\bibitem [{\citenamefont {A\~na\ nos}\ \emph {et~al.}(2003)\citenamefont
  {A\~na\ nos}, \citenamefont {Camblong},\ and\ \citenamefont
  {Ord\'o\~nez}}]{PhysRevD.68.025006}%
  \BibitemOpen
  \bibfield  {author} {\bibinfo {author} {\bibfnamefont {G.~N.~J.}\
  \bibnamefont {A\~na\ nos}}, \bibinfo {author} {\bibfnamefont {H.~E.}\
  \bibnamefont {Camblong}}, \ and\ \bibinfo {author} {\bibfnamefont {C.~R.}\
  \bibnamefont {Ord\'o\~nez}},\ }\href {\doibase 10.1103/PhysRevD.68.025006}
  {\bibfield  {journal} {\bibinfo  {journal} {Phys. Rev. D}\ }\textbf {\bibinfo
  {volume} {68}},\ \bibinfo {pages} {025006} (\bibinfo {year}
  {2003})}\BibitemShut {NoStop}%
\bibitem [{\citenamefont {Hammer}\ and\ \citenamefont
  {Swingle}(2006)}]{Hammer:2005sa}%
  \BibitemOpen
  \bibfield  {author} {\bibinfo {author} {\bibfnamefont {H.~W.}\ \bibnamefont
  {Hammer}}\ and\ \bibinfo {author} {\bibfnamefont {B.~G.}\ \bibnamefont
  {Swingle}},\ }\href {\doibase 10.1016/j.aop.2005.04.017} {\bibfield
  {journal} {\bibinfo  {journal} {Annals Phys.}\ }\textbf {\bibinfo {volume}
  {321}},\ \bibinfo {pages} {306} (\bibinfo {year} {2006})},\ \Eprint
  {http://arxiv.org/abs/quant-ph/0503074} {arXiv:quant-ph/0503074 [quant-ph]}
  \BibitemShut {NoStop}%
%%CITATION = QUANT-PH/0503074;%%
\bibitem [{\citenamefont {Braaten}\ and\ \citenamefont
  {Phillips}(2004)}]{Braaten2004}%
  \BibitemOpen
  \bibfield  {author} {\bibinfo {author} {\bibfnamefont {E.}~\bibnamefont
  {Braaten}}\ and\ \bibinfo {author} {\bibfnamefont {D.}~\bibnamefont
  {Phillips}},\ }\href {\doibase 10.1103/PhysRevA.70.052111} {\bibfield
  {journal} {\bibinfo  {journal} {Phys. Rev. A}\ }\textbf {\bibinfo {volume}
  {70}},\ \bibinfo {pages} {052111} (\bibinfo {year} {2004})}\BibitemShut
  {NoStop}%
\bibitem [{\citenamefont {Kaplan}\ \emph {et~al.}(2009)\citenamefont {Kaplan},
  \citenamefont {Lee}, \citenamefont {Son},\ and\ \citenamefont
  {Stephanov}}]{Kaplan:2009kr}%
  \BibitemOpen
  \bibfield  {author} {\bibinfo {author} {\bibfnamefont {D.~B.}\ \bibnamefont
  {Kaplan}}, \bibinfo {author} {\bibfnamefont {J.-W.}\ \bibnamefont {Lee}},
  \bibinfo {author} {\bibfnamefont {D.~T.}\ \bibnamefont {Son}}, \ and\
  \bibinfo {author} {\bibfnamefont {M.~A.}\ \bibnamefont {Stephanov}},\ }\href
  {\doibase 10.1103/PhysRevD.80.125005} {\bibfield  {journal} {\bibinfo
  {journal} {Phys. Rev. D}\ }\textbf {\bibinfo {volume} {80}},\ \bibinfo
  {pages} {125005} (\bibinfo {year} {2009})}\BibitemShut {NoStop}%
\bibitem [{\citenamefont {Efimov}(1970)}]{efimov1970energy}%
  \BibitemOpen
  \bibfield  {author} {\bibinfo {author} {\bibfnamefont {V.}~\bibnamefont
  {Efimov}},\ }\href {\doibase 10.1016/0370-2693(70)90349-7} {\bibfield
  {journal} {\bibinfo  {journal} {Physics Letters B}\ }\textbf {\bibinfo
  {volume} {33}},\ \bibinfo {pages} {563} (\bibinfo {year} {1970})}\BibitemShut
  {NoStop}%
\bibitem [{\citenamefont {Efimov}(1971)}]{Efimov1971}%
  \BibitemOpen
  \bibfield  {author} {\bibinfo {author} {\bibfnamefont {V.}~\bibnamefont
  {Efimov}},\ }\href
  {https://www.uibk.ac.at/exphys/ultracold/projects/levt/FourBodies/SovJNucPhys12.589.efimov.pdf}
  {\bibfield  {journal} {\bibinfo  {journal} {Sov. J. Nucl. Phys}\ }\textbf
  {\bibinfo {volume} {12}},\ \bibinfo {pages} {589} (\bibinfo {year}
  {1971})}\BibitemShut {NoStop}%
\bibitem [{\citenamefont {Braaten}\ and\ \citenamefont
  {Hammer}(2006)}]{Braaten2006259}%
  \BibitemOpen
  \bibfield  {author} {\bibinfo {author} {\bibfnamefont {E.}~\bibnamefont
  {Braaten}}\ and\ \bibinfo {author} {\bibfnamefont {H.-W.}\ \bibnamefont
  {Hammer}},\ }\href {\doibase http://dx.doi.org/10.1016/j.physrep.2006.03.001}
  {\bibfield  {journal} {\bibinfo  {journal} {Physics Reports}\ }\textbf
  {\bibinfo {volume} {428}},\ \bibinfo {pages} {259 } (\bibinfo {year}
  {2006})}\BibitemShut {NoStop}%
\bibitem [{\citenamefont {L\'evy-Leblond}(1967)}]{levy1967electron}%
  \BibitemOpen
  \bibfield  {author} {\bibinfo {author} {\bibfnamefont {J.-M.}\ \bibnamefont
  {L\'evy-Leblond}},\ }\href {\doibase 10.1103/PhysRev.153.1} {\bibfield
  {journal} {\bibinfo  {journal} {Phys. Rev.}\ }\textbf {\bibinfo {volume}
  {153}},\ \bibinfo {pages} {1} (\bibinfo {year} {1967})}\BibitemShut {NoStop}%
\bibitem [{\citenamefont {Kolomeisky}\ and\ \citenamefont
  {Straley}(1992)}]{PhysRevB.46.12664}%
  \BibitemOpen
  \bibfield  {author} {\bibinfo {author} {\bibfnamefont {E.~B.}\ \bibnamefont
  {Kolomeisky}}\ and\ \bibinfo {author} {\bibfnamefont {J.~P.}\ \bibnamefont
  {Straley}},\ }\href {\doibase 10.1103/PhysRevB.46.12664} {\bibfield
  {journal} {\bibinfo  {journal} {Phys. Rev. B}\ }\textbf {\bibinfo {volume}
  {46}},\ \bibinfo {pages} {12664} (\bibinfo {year} {1992})}\BibitemShut
  {NoStop}%
\bibitem [{\citenamefont {Gupta}\ and\ \citenamefont
  {Rajeev}(1993)}]{PhysRevD.48.5940}%
  \BibitemOpen
  \bibfield  {author} {\bibinfo {author} {\bibfnamefont {K.~S.}\ \bibnamefont
  {Gupta}}\ and\ \bibinfo {author} {\bibfnamefont {S.~G.}\ \bibnamefont
  {Rajeev}},\ }\href {\doibase 10.1103/PhysRevD.48.5940} {\bibfield  {journal}
  {\bibinfo  {journal} {Phys. Rev. D}\ }\textbf {\bibinfo {volume} {48}},\
  \bibinfo {pages} {5940} (\bibinfo {year} {1993})}\BibitemShut {NoStop}%
\bibitem [{\citenamefont {Camblong}\ \emph {et~al.}(2001)\citenamefont
  {Camblong}, \citenamefont {Epele}, \citenamefont {Fanchiotti},\ and\
  \citenamefont {Garc\'{\i}a~Canal}}]{CamblongEpeleFanchiottiEtAl2001}%
  \BibitemOpen
  \bibfield  {author} {\bibinfo {author} {\bibfnamefont {H.~E.}\ \bibnamefont
  {Camblong}}, \bibinfo {author} {\bibfnamefont {L.~N.}\ \bibnamefont {Epele}},
  \bibinfo {author} {\bibfnamefont {H.}~\bibnamefont {Fanchiotti}}, \ and\
  \bibinfo {author} {\bibfnamefont {C.~A.}\ \bibnamefont {Garc\'{\i}a~Canal}},\
  }\href {\doibase 10.1103/PhysRevLett.87.220402} {\bibfield  {journal}
  {\bibinfo  {journal} {Phys. Rev. Lett.}\ }\textbf {\bibinfo {volume} {87}},\
  \bibinfo {pages} {220402} (\bibinfo {year} {2001})}\BibitemShut {NoStop}%
\bibitem [{\citenamefont {Nisoli}\ and\ \citenamefont
  {Bishop}(2014)}]{nisoli2014attractive}%
  \BibitemOpen
  \bibfield  {author} {\bibinfo {author} {\bibfnamefont {C.}~\bibnamefont
  {Nisoli}}\ and\ \bibinfo {author} {\bibfnamefont {A.~R.}\ \bibnamefont
  {Bishop}},\ }\href {\doibase 10.1103/PhysRevLett.112.070401} {\bibfield
  {journal} {\bibinfo  {journal} {Phys. Rev. Lett.}\ }\textbf {\bibinfo
  {volume} {112}},\ \bibinfo {pages} {070401} (\bibinfo {year}
  {2014})}\BibitemShut {NoStop}%
\bibitem [{\citenamefont {Govindarajan}\ \emph {et~al.}(2000)\citenamefont
  {Govindarajan}, \citenamefont {Suneeta},\ and\ \citenamefont
  {Vaidya}}]{Govindarajan:2000ag}%
  \BibitemOpen
  \bibfield  {author} {\bibinfo {author} {\bibfnamefont {T.~R.}\ \bibnamefont
  {Govindarajan}}, \bibinfo {author} {\bibfnamefont {V.}~\bibnamefont
  {Suneeta}}, \ and\ \bibinfo {author} {\bibfnamefont {S.}~\bibnamefont
  {Vaidya}},\ }\href {\doibase 10.1016/S0550-3213(00)00336-9} {\bibfield
  {journal} {\bibinfo  {journal} {Nucl. Phys.}\ }\textbf {\bibinfo {volume}
  {B583}},\ \bibinfo {pages} {291} (\bibinfo {year} {2000})},\ \Eprint
  {http://arxiv.org/abs/hep-th/0002036} {arXiv:hep-th/0002036 [hep-th]}
  \BibitemShut {NoStop}%
%%CITATION = HEP-TH/0002036;%%
\bibitem [{\citenamefont {Camblong}\ and\ \citenamefont
  {Ordonez}(2003)}]{Camblong:2003mz}%
  \BibitemOpen
  \bibfield  {author} {\bibinfo {author} {\bibfnamefont {H.~E.}\ \bibnamefont
  {Camblong}}\ and\ \bibinfo {author} {\bibfnamefont {C.~R.}\ \bibnamefont
  {Ordonez}},\ }\href {\doibase 10.1103/PhysRevD.68.125013} {\bibfield
  {journal} {\bibinfo  {journal} {Phys. Rev.}\ }\textbf {\bibinfo {volume}
  {D68}},\ \bibinfo {pages} {125013} (\bibinfo {year} {2003})},\ \Eprint
  {http://arxiv.org/abs/hep-th/0303166} {arXiv:hep-th/0303166 [hep-th]}
  \BibitemShut {NoStop}%
%%CITATION = HEP-TH/0303166;%%
\bibitem [{\citenamefont {Bellucci}\ \emph {et~al.}(2003)\citenamefont
  {Bellucci}, \citenamefont {Galajinsky}, \citenamefont {Ivanov},\ and\
  \citenamefont {Krivonos}}]{Bellucci200399}%
  \BibitemOpen
  \bibfield  {author} {\bibinfo {author} {\bibfnamefont {S.}~\bibnamefont
  {Bellucci}}, \bibinfo {author} {\bibfnamefont {A.}~\bibnamefont
  {Galajinsky}}, \bibinfo {author} {\bibfnamefont {E.}~\bibnamefont {Ivanov}},
  \ and\ \bibinfo {author} {\bibfnamefont {S.}~\bibnamefont {Krivonos}},\
  }\href {\doibase https://doi.org/10.1016/S0370-2693(03)00040-6} {\bibfield
  {journal} {\bibinfo  {journal} {Physics Letters B}\ }\textbf {\bibinfo
  {volume} {555}},\ \bibinfo {pages} {99 } (\bibinfo {year}
  {2003})}\BibitemShut {NoStop}%
\bibitem [{\citenamefont {Jackiw}(1995)}]{jackiw1995diverse}%
  \BibitemOpen
  \bibfield  {author} {\bibinfo {author} {\bibfnamefont {R.~W.}\ \bibnamefont
  {Jackiw}},\ }\href@noop {} {\emph {\bibinfo {title} {Diverse topics in
  theoretical and mathematical physics}}}\ (\bibinfo  {publisher} {World
  Scientific},\ \bibinfo {year} {1995})\BibitemShut {NoStop}%
\bibitem [{Note1()}]{Note1}%
  \BibitemOpen
  \bibinfo {note} {Thus, for $d=2$, an attractive potential is always
  overcritical.}\BibitemShut {Stop}%
\bibitem [{\citenamefont {Jensen}\ \emph {et~al.}(2010)\citenamefont {Jensen},
  \citenamefont {Karch}, \citenamefont {Son},\ and\ \citenamefont
  {Thompson}}]{Jensen:2010ga}%
  \BibitemOpen
  \bibfield  {author} {\bibinfo {author} {\bibfnamefont {K.}~\bibnamefont
  {Jensen}}, \bibinfo {author} {\bibfnamefont {A.}~\bibnamefont {Karch}},
  \bibinfo {author} {\bibfnamefont {D.~T.}\ \bibnamefont {Son}}, \ and\
  \bibinfo {author} {\bibfnamefont {E.~G.}\ \bibnamefont {Thompson}},\ }\href
  {\doibase 10.1103/PhysRevLett.105.041601} {\bibfield  {journal} {\bibinfo
  {journal} {Phys. Rev. Lett.}\ }\textbf {\bibinfo {volume} {105}},\ \bibinfo
  {pages} {041601} (\bibinfo {year} {2010})},\ \Eprint
  {http://arxiv.org/abs/1002.3159} {arXiv:1002.3159 [hep-th]} \BibitemShut
  {NoStop}%
%%CITATION = ARXIV:1002.3159;%%
\bibitem [{\citenamefont {Jensen}(2010)}]{Jensen:2010vx}%
  \BibitemOpen
  \bibfield  {author} {\bibinfo {author} {\bibfnamefont {K.}~\bibnamefont
  {Jensen}},\ }\href {\doibase 10.1103/PhysRevD.82.046005} {\bibfield
  {journal} {\bibinfo  {journal} {Phys. Rev.}\ }\textbf {\bibinfo {volume}
  {D82}},\ \bibinfo {pages} {046005} (\bibinfo {year} {2010})},\ \Eprint
  {http://arxiv.org/abs/1006.3066} {arXiv:1006.3066 [hep-th]} \BibitemShut
  {NoStop}%
%%CITATION = ARXIV:1006.3066;%%
\bibitem [{\citenamefont {Jensen}(2011)}]{Jensen:2011af}%
  \BibitemOpen
  \bibfield  {author} {\bibinfo {author} {\bibfnamefont {K.}~\bibnamefont
  {Jensen}},\ }\href {\doibase 10.1103/PhysRevLett.107.231601} {\bibfield
  {journal} {\bibinfo  {journal} {Phys. Rev. Lett.}\ }\textbf {\bibinfo
  {volume} {107}},\ \bibinfo {pages} {231601} (\bibinfo {year} {2011})},\
  \Eprint {http://arxiv.org/abs/1108.0421} {arXiv:1108.0421 [hep-th]}
  \BibitemShut {NoStop}%
%%CITATION = ARXIV:1108.0421;%%
\bibitem [{\citenamefont {{Derrida}}\ and\ \citenamefont
  {{Retaux}}(2014)}]{2014JSP...156..268D}%
  \BibitemOpen
  \bibfield  {author} {\bibinfo {author} {\bibfnamefont {B.}~\bibnamefont
  {{Derrida}}}\ and\ \bibinfo {author} {\bibfnamefont {M.}~\bibnamefont
  {{Retaux}}},\ }\href {\doibase 10.1007/s10955-014-1006-y} {\bibfield
  {journal} {\bibinfo  {journal} {Journal of Statistical Physics}\ }\textbf
  {\bibinfo {volume} {156}},\ \bibinfo {pages} {268} (\bibinfo {year}
  {2014})},\ \Eprint {http://arxiv.org/abs/1401.6919} {arXiv:1401.6919
  [cond-mat.stat-mech]} \BibitemShut {NoStop}%
\bibitem [{\citenamefont {Gies}\ and\ \citenamefont
  {Torgrimsson}(2016)}]{Gies:2015hia}%
  \BibitemOpen
  \bibfield  {author} {\bibinfo {author} {\bibfnamefont {H.}~\bibnamefont
  {Gies}}\ and\ \bibinfo {author} {\bibfnamefont {G.}~\bibnamefont
  {Torgrimsson}},\ }\href {\doibase 10.1103/PhysRevLett.116.090406} {\bibfield
  {journal} {\bibinfo  {journal} {Phys. Rev. Lett.}\ }\textbf {\bibinfo
  {volume} {116}},\ \bibinfo {pages} {090406} (\bibinfo {year} {2016})},\
  \Eprint {http://arxiv.org/abs/1507.07802} {arXiv:1507.07802 [hep-ph]}
  \BibitemShut {NoStop}%
%%CITATION = ARXIV:1507.07802;%%
\bibitem [{\citenamefont {Ovdat}\ \emph {et~al.}(2017)\citenamefont {Ovdat},
  \citenamefont {Mao}, \citenamefont {Jiang}, \citenamefont {Andrei},\ and\
  \citenamefont {Akkermans}}]{OvdatMaoJiangEtAl2017}%
  \BibitemOpen
  \bibfield  {author} {\bibinfo {author} {\bibfnamefont {O.}~\bibnamefont
  {Ovdat}}, \bibinfo {author} {\bibfnamefont {J.}~\bibnamefont {Mao}}, \bibinfo
  {author} {\bibfnamefont {Y.}~\bibnamefont {Jiang}}, \bibinfo {author}
  {\bibfnamefont {E.~Y.}\ \bibnamefont {Andrei}}, \ and\ \bibinfo {author}
  {\bibfnamefont {E.}~\bibnamefont {Akkermans}},\ }\href {\doibase
  10.1038/s41467-017-00591-8} {\bibfield  {journal} {\bibinfo  {journal}
  {Nature Communications}\ }\textbf {\bibinfo {volume} {8}},\ \bibinfo {pages}
  {507} (\bibinfo {year} {2017})}\BibitemShut {NoStop}%
\bibitem [{\citenamefont {Alexandre}(2011)}]{Alexandre:2011kr}%
  \BibitemOpen
  \bibfield  {author} {\bibinfo {author} {\bibfnamefont {J.}~\bibnamefont
  {Alexandre}},\ }\href {\doibase 10.1142/S0217751X11054656} {\bibfield
  {journal} {\bibinfo  {journal} {Int. J. Mod. Phys.}\ }\textbf {\bibinfo
  {volume} {A26}},\ \bibinfo {pages} {4523} (\bibinfo {year} {2011})},\ \Eprint
  {http://arxiv.org/abs/1109.5629} {arXiv:1109.5629 [hep-ph]} \BibitemShut
  {NoStop}%
%%CITATION = ARXIV:1109.5629;%%
\bibitem [{\citenamefont {Hornreich}\ \emph {et~al.}(1975)\citenamefont
  {Hornreich}, \citenamefont {Luban},\ and\ \citenamefont
  {Shtrikman}}]{PhysRevLett.35.1678}%
  \BibitemOpen
  \bibfield  {author} {\bibinfo {author} {\bibfnamefont {R.~M.}\ \bibnamefont
  {Hornreich}}, \bibinfo {author} {\bibfnamefont {M.}~\bibnamefont {Luban}}, \
  and\ \bibinfo {author} {\bibfnamefont {S.}~\bibnamefont {Shtrikman}},\ }\href
  {\doibase 10.1103/PhysRevLett.35.1678} {\bibfield  {journal} {\bibinfo
  {journal} {Phys. Rev. Lett.}\ }\textbf {\bibinfo {volume} {35}},\ \bibinfo
  {pages} {1678} (\bibinfo {year} {1975})}\BibitemShut {NoStop}%
\bibitem [{\citenamefont {Grinstein}(1981)}]{PhysRevB.23.4615}%
  \BibitemOpen
  \bibfield  {author} {\bibinfo {author} {\bibfnamefont {G.}~\bibnamefont
  {Grinstein}},\ }\href {\doibase 10.1103/PhysRevB.23.4615} {\bibfield
  {journal} {\bibinfo  {journal} {Phys. Rev. B}\ }\textbf {\bibinfo {volume}
  {23}},\ \bibinfo {pages} {4615} (\bibinfo {year} {1981})}\BibitemShut
  {NoStop}%
\bibitem [{\citenamefont {Fradkin}\ \emph {et~al.}(2004)\citenamefont
  {Fradkin}, \citenamefont {Huse}, \citenamefont {Moessner}, \citenamefont
  {Oganesyan},\ and\ \citenamefont {Sondhi}}]{PhysRevB.69.224415}%
  \BibitemOpen
  \bibfield  {author} {\bibinfo {author} {\bibfnamefont {E.}~\bibnamefont
  {Fradkin}}, \bibinfo {author} {\bibfnamefont {D.~A.}\ \bibnamefont {Huse}},
  \bibinfo {author} {\bibfnamefont {R.}~\bibnamefont {Moessner}}, \bibinfo
  {author} {\bibfnamefont {V.}~\bibnamefont {Oganesyan}}, \ and\ \bibinfo
  {author} {\bibfnamefont {S.~L.}\ \bibnamefont {Sondhi}},\ }\href {\doibase
  10.1103/PhysRevB.69.224415} {\bibfield  {journal} {\bibinfo  {journal} {Phys.
  Rev. B}\ }\textbf {\bibinfo {volume} {69}},\ \bibinfo {pages} {224415}
  (\bibinfo {year} {2004})}\BibitemShut {NoStop}%
\bibitem [{\citenamefont {Vishwanath}\ \emph {et~al.}(2004)\citenamefont
  {Vishwanath}, \citenamefont {Balents},\ and\ \citenamefont
  {Senthil}}]{PhysRevB.69.224416}%
  \BibitemOpen
  \bibfield  {author} {\bibinfo {author} {\bibfnamefont {A.}~\bibnamefont
  {Vishwanath}}, \bibinfo {author} {\bibfnamefont {L.}~\bibnamefont {Balents}},
  \ and\ \bibinfo {author} {\bibfnamefont {T.}~\bibnamefont {Senthil}},\ }\href
  {\doibase 10.1103/PhysRevB.69.224416} {\bibfield  {journal} {\bibinfo
  {journal} {Phys. Rev. B}\ }\textbf {\bibinfo {volume} {69}},\ \bibinfo
  {pages} {224416} (\bibinfo {year} {2004})}\BibitemShut {NoStop}%
\bibitem [{\citenamefont {Ardonne}\ \emph {et~al.}(2004)\citenamefont
  {Ardonne}, \citenamefont {Fendley},\ and\ \citenamefont
  {Fradkin}}]{Ardonne:2003wa}%
  \BibitemOpen
  \bibfield  {author} {\bibinfo {author} {\bibfnamefont {E.}~\bibnamefont
  {Ardonne}}, \bibinfo {author} {\bibfnamefont {P.}~\bibnamefont {Fendley}}, \
  and\ \bibinfo {author} {\bibfnamefont {E.}~\bibnamefont {Fradkin}},\ }\href
  {\doibase 10.1016/j.aop.2004.01.004} {\bibfield  {journal} {\bibinfo
  {journal} {Annals Phys.}\ }\textbf {\bibinfo {volume} {310}},\ \bibinfo
  {pages} {493} (\bibinfo {year} {2004})},\ \Eprint
  {http://arxiv.org/abs/cond-mat/0311466} {arXiv:cond-mat/0311466 [cond-mat]}
  \BibitemShut {NoStop}%
%%CITATION = COND-MAT/0311466;%%
\bibitem [{\citenamefont {Mukohyama}(2010)}]{Mukohyama:2010xz}%
  \BibitemOpen
  \bibfield  {author} {\bibinfo {author} {\bibfnamefont {S.}~\bibnamefont
  {Mukohyama}},\ }\href {\doibase 10.1088/0264-9381/27/22/223101} {\bibfield
  {journal} {\bibinfo  {journal} {Class. Quant. Grav.}\ }\textbf {\bibinfo
  {volume} {27}},\ \bibinfo {pages} {223101} (\bibinfo {year} {2010})},\
  \Eprint {http://arxiv.org/abs/1007.5199} {arXiv:1007.5199 [hep-th]}
  \BibitemShut {NoStop}%
%%CITATION = ARXIV:1007.5199;%%
\bibitem [{\citenamefont {Reuter}(1998)}]{PhysRevD.57.971}%
  \BibitemOpen
  \bibfield  {author} {\bibinfo {author} {\bibfnamefont {M.}~\bibnamefont
  {Reuter}},\ }\href {\doibase 10.1103/PhysRevD.57.971} {\bibfield  {journal}
  {\bibinfo  {journal} {Phys. Rev. D}\ }\textbf {\bibinfo {volume} {57}},\
  \bibinfo {pages} {971} (\bibinfo {year} {1998})}\BibitemShut {NoStop}%
\bibitem [{\citenamefont {Kachru}\ \emph {et~al.}(2008)\citenamefont {Kachru},
  \citenamefont {Liu},\ and\ \citenamefont {Mulligan}}]{Kachru:2008yh}%
  \BibitemOpen
  \bibfield  {author} {\bibinfo {author} {\bibfnamefont {S.}~\bibnamefont
  {Kachru}}, \bibinfo {author} {\bibfnamefont {X.}~\bibnamefont {Liu}}, \ and\
  \bibinfo {author} {\bibfnamefont {M.}~\bibnamefont {Mulligan}},\ }\href
  {\doibase 10.1103/PhysRevD.78.106005} {\bibfield  {journal} {\bibinfo
  {journal} {Phys. Rev.}\ }\textbf {\bibinfo {volume} {D78}},\ \bibinfo {pages}
  {106005} (\bibinfo {year} {2008})},\ \Eprint {http://arxiv.org/abs/0808.1725}
  {arXiv:0808.1725 [hep-th]} \BibitemShut {NoStop}%
%%CITATION = ARXIV:0808.1725;%%
\bibitem [{\citenamefont {Horava}(2009{\natexlab{a}})}]{Horava:2009if}%
  \BibitemOpen
  \bibfield  {author} {\bibinfo {author} {\bibfnamefont {P.}~\bibnamefont
  {Horava}},\ }\href {\doibase 10.1103/PhysRevLett.102.161301} {\bibfield
  {journal} {\bibinfo  {journal} {Phys. Rev. Lett.}\ }\textbf {\bibinfo
  {volume} {102}},\ \bibinfo {pages} {161301} (\bibinfo {year}
  {2009}{\natexlab{a}})},\ \Eprint {http://arxiv.org/abs/0902.3657}
  {arXiv:0902.3657 [hep-th]} \BibitemShut {NoStop}%
%%CITATION = ARXIV:0902.3657;%%
\bibitem [{\citenamefont {Horava}(2009{\natexlab{b}})}]{Horava:2009uw}%
  \BibitemOpen
  \bibfield  {author} {\bibinfo {author} {\bibfnamefont {P.}~\bibnamefont
  {Horava}},\ }\href {\doibase 10.1103/PhysRevD.79.084008} {\bibfield
  {journal} {\bibinfo  {journal} {Phys. Rev.}\ }\textbf {\bibinfo {volume}
  {D79}},\ \bibinfo {pages} {084008} (\bibinfo {year} {2009}{\natexlab{b}})},\
  \Eprint {http://arxiv.org/abs/0901.3775} {arXiv:0901.3775 [hep-th]}
  \BibitemShut {NoStop}%
%%CITATION = ARXIV:0901.3775;%%
\bibitem [{\citenamefont {Gies}\ \emph {et~al.}(2016)\citenamefont {Gies},
  \citenamefont {Knorr}, \citenamefont {Lippoldt},\ and\ \citenamefont
  {Saueressig}}]{Gies:2016con}%
  \BibitemOpen
  \bibfield  {author} {\bibinfo {author} {\bibfnamefont {H.}~\bibnamefont
  {Gies}}, \bibinfo {author} {\bibfnamefont {B.}~\bibnamefont {Knorr}},
  \bibinfo {author} {\bibfnamefont {S.}~\bibnamefont {Lippoldt}}, \ and\
  \bibinfo {author} {\bibfnamefont {F.}~\bibnamefont {Saueressig}},\ }\href
  {\doibase 10.1103/PhysRevLett.116.211302} {\bibfield  {journal} {\bibinfo
  {journal} {Phys. Rev. Lett.}\ }\textbf {\bibinfo {volume} {116}},\ \bibinfo
  {pages} {211302} (\bibinfo {year} {2016})},\ \Eprint
  {http://arxiv.org/abs/1601.01800} {arXiv:1601.01800 [hep-th]} \BibitemShut
  {NoStop}%
%%CITATION = ARXIV:1601.01800;%%
\bibitem [{\citenamefont {Brauner}(2010)}]{Brauner:2010wm}%
  \BibitemOpen
  \bibfield  {author} {\bibinfo {author} {\bibfnamefont {T.}~\bibnamefont
  {Brauner}},\ }\href {\doibase 10.3390/sym2020609} {\bibfield  {journal}
  {\bibinfo  {journal} {Symmetry}\ }\textbf {\bibinfo {volume} {2}},\ \bibinfo
  {pages} {609} (\bibinfo {year} {2010})},\ \Eprint
  {http://arxiv.org/abs/1001.5212} {arXiv:1001.5212 [hep-th]} \BibitemShut
  {NoStop}%
%%CITATION = ARXIV:1001.5212;%%
\bibitem [{\citenamefont {Horava}(2011)}]{Horava:2008jf}%
  \BibitemOpen
  \bibfield  {author} {\bibinfo {author} {\bibfnamefont {P.}~\bibnamefont
  {Horava}},\ }\href {\doibase 10.1016/j.physletb.2010.09.055} {\bibfield
  {journal} {\bibinfo  {journal} {Phys. Lett.}\ }\textbf {\bibinfo {volume}
  {B694}},\ \bibinfo {pages} {172} (\bibinfo {year} {2011})},\ \Eprint
  {http://arxiv.org/abs/0811.2217} {arXiv:0811.2217 [hep-th]} \BibitemShut
  {NoStop}%
%%CITATION = ARXIV:0811.2217;%%
\bibitem [{\citenamefont {Das}\ and\ \citenamefont
  {Murthy}(2010)}]{Das:2009fb}%
  \BibitemOpen
  \bibfield  {author} {\bibinfo {author} {\bibfnamefont {S.~R.}\ \bibnamefont
  {Das}}\ and\ \bibinfo {author} {\bibfnamefont {G.}~\bibnamefont {Murthy}},\
  }\href {\doibase 10.1103/PhysRevLett.104.181601} {\bibfield  {journal}
  {\bibinfo  {journal} {Phys. Rev. Lett.}\ }\textbf {\bibinfo {volume} {104}},\
  \bibinfo {pages} {181601} (\bibinfo {year} {2010})},\ \Eprint
  {http://arxiv.org/abs/0909.3064} {arXiv:0909.3064 [hep-th]} \BibitemShut
  {NoStop}%
%%CITATION = ARXIV:0909.3064;%%
\bibitem [{\citenamefont {Farias}\ \emph {et~al.}(2012)\citenamefont {Farias},
  \citenamefont {Gomes}, \citenamefont {Nascimento}, \citenamefont {Petrov},\
  and\ \citenamefont {da~Silva}}]{Farias:2011aa}%
  \BibitemOpen
  \bibfield  {author} {\bibinfo {author} {\bibfnamefont {C.~F.}\ \bibnamefont
  {Farias}}, \bibinfo {author} {\bibfnamefont {M.}~\bibnamefont {Gomes}},
  \bibinfo {author} {\bibfnamefont {J.~R.}\ \bibnamefont {Nascimento}},
  \bibinfo {author} {\bibfnamefont {A.~{\relax Yu}.}\ \bibnamefont {Petrov}}, \
  and\ \bibinfo {author} {\bibfnamefont {A.~J.}\ \bibnamefont {da~Silva}},\
  }\href {\doibase 10.1103/PhysRevD.85.127701} {\bibfield  {journal} {\bibinfo
  {journal} {Phys. Rev.}\ }\textbf {\bibinfo {volume} {D85}},\ \bibinfo {pages}
  {127701} (\bibinfo {year} {2012})},\ \Eprint {http://arxiv.org/abs/1112.2081}
  {arXiv:1112.2081 [hep-th]} \BibitemShut {NoStop}%
%%CITATION = ARXIV:1112.2081;%%
\bibitem [{\citenamefont {Pal}\ and\ \citenamefont
  {Grinstein}(2016)}]{Pal:2016rpz}%
  \BibitemOpen
  \bibfield  {author} {\bibinfo {author} {\bibfnamefont {S.}~\bibnamefont
  {Pal}}\ and\ \bibinfo {author} {\bibfnamefont {B.}~\bibnamefont
  {Grinstein}},\ }\href {\doibase 10.1007/JHEP12(2016)012} {\bibfield
  {journal} {\bibinfo  {journal} {JHEP}\ }\textbf {\bibinfo {volume} {12}},\
  \bibinfo {pages} {012} (\bibinfo {year} {2016})},\ \Eprint
  {http://arxiv.org/abs/1605.02748} {arXiv:1605.02748 [hep-th]} \BibitemShut
  {NoStop}%
%%CITATION = ARXIV:1605.02748;%%
\bibitem [{Note2()}]{Note2}%
  \BibitemOpen
  \bibinfo {note} {Additional boundary terms will be required to make the
  variation of the action vanish on boundary conditions that make the
  corresponding Hamiltonian operator self-adjoint \cite
  {Asorey:2006pr}.}\BibitemShut {Stop}%
\bibitem [{\citenamefont {Bonneau}\ \emph {et~al.}(2001)\citenamefont
  {Bonneau}, \citenamefont {Faraut},\ and\ \citenamefont
  {Valent}}]{Bonneau:1999zq}%
  \BibitemOpen
  \bibfield  {author} {\bibinfo {author} {\bibfnamefont {G.}~\bibnamefont
  {Bonneau}}, \bibinfo {author} {\bibfnamefont {J.}~\bibnamefont {Faraut}}, \
  and\ \bibinfo {author} {\bibfnamefont {G.}~\bibnamefont {Valent}},\ }\href
  {\doibase 10.1119/1.1328351} {\bibfield  {journal} {\bibinfo  {journal} {Am.
  J. Phys.}\ }\textbf {\bibinfo {volume} {69}},\ \bibinfo {pages} {322}
  (\bibinfo {year} {2001})},\ \Eprint {http://arxiv.org/abs/quant-ph/0103153}
  {arXiv:quant-ph/0103153 [quant-ph]} \BibitemShut {NoStop}%
%%CITATION = QUANT-PH/0103153;%%
\bibitem [{\citenamefont {{Ibort}}\ \emph {et~al.}(2015)\citenamefont
  {{Ibort}}, \citenamefont {{Lled{\'o}}},\ and\ \citenamefont
  {{P{\'e}rez-Pardo}}}]{2015AnHP...16.2367I}%
  \BibitemOpen
  \bibfield  {author} {\bibinfo {author} {\bibfnamefont {A.}~\bibnamefont
  {{Ibort}}}, \bibinfo {author} {\bibfnamefont {F.}~\bibnamefont
  {{Lled{\'o}}}}, \ and\ \bibinfo {author} {\bibfnamefont {J.~M.}\ \bibnamefont
  {{P{\'e}rez-Pardo}}},\ }\href {\doibase 10.1007/s00023-014-0379-4} {\bibfield
   {journal} {\bibinfo  {journal} {Annales Henri Poincar{\'e}}\ }\textbf
  {\bibinfo {volume} {16}},\ \bibinfo {pages} {2367} (\bibinfo {year}
  {2015})},\ \Eprint {http://arxiv.org/abs/1402.5537} {arXiv:1402.5537
  [math-ph]} \BibitemShut {NoStop}%
\bibitem [{\citenamefont {Gitman}\ \emph {et~al.}(2009)\citenamefont {Gitman},
  \citenamefont {Tyutin},\ and\ \citenamefont {Voronov}}]{Gitman:2009era}%
  \BibitemOpen
  \bibfield  {author} {\bibinfo {author} {\bibfnamefont {D.~M.}\ \bibnamefont
  {Gitman}}, \bibinfo {author} {\bibfnamefont {I.~V.}\ \bibnamefont {Tyutin}},
  \ and\ \bibinfo {author} {\bibfnamefont {B.~L.}\ \bibnamefont {Voronov}},\
  }\href {\doibase 10.1088/1751-8113/43/14/145205} {\  (\bibinfo {year}
  {2009}),\ 10.1088/1751-8113/43/14/145205},\ \Eprint
  {http://arxiv.org/abs/0903.5277} {arXiv:0903.5277 [quant-ph]} \BibitemShut
  {NoStop}%
%%CITATION = ARXIV:0903.5277;%%
\bibitem [{\citenamefont {Meetz}(1964)}]{Meetz1964}%
  \BibitemOpen
  \bibfield  {author} {\bibinfo {author} {\bibfnamefont {K.}~\bibnamefont
  {Meetz}},\ }\href {\doibase 10.1007/BF02750010} {\bibfield  {journal}
  {\bibinfo  {journal} {Il Nuovo Cimento (1955-1965)}\ }\textbf {\bibinfo
  {volume} {34}},\ \bibinfo {pages} {690} (\bibinfo {year} {1964})}\BibitemShut
  {NoStop}%
\bibitem [{\citenamefont {Gitman}\ \emph
  {et~al.}(2012{\natexlab{a}})\citenamefont {Gitman}, \citenamefont {Tyutin},\
  and\ \citenamefont {Voronov}}]{gitman2012self}%
  \BibitemOpen
  \bibfield  {author} {\bibinfo {author} {\bibfnamefont {D.~M.}\ \bibnamefont
  {Gitman}}, \bibinfo {author} {\bibfnamefont {I.}~\bibnamefont {Tyutin}}, \
  and\ \bibinfo {author} {\bibfnamefont {B.~L.}\ \bibnamefont {Voronov}},\
  }\href@noop {} {\emph {\bibinfo {title} {Self-adjoint Extensions in Quantum
  Mechanics: General Theory and Applications to Schr{\"o}dinger and Dirac
  Equations with Singular Potentials}}},\ Vol.~\bibinfo {volume} {62}\
  (\bibinfo  {publisher} {Springer},\ \bibinfo {year} {2012})\BibitemShut
  {NoStop}%
\bibitem [{\citenamefont {Luke}(1969)}]{luke1969special}%
  \BibitemOpen
  \bibfield  {author} {\bibinfo {author} {\bibfnamefont {Y.}~\bibnamefont
  {Luke}},\ }\href {https://books.google.com/books?id=huuO6mKbVoEC} {\emph
  {\bibinfo {title} {The Special Functions and Their Approximations}}},\
  Mathematics in Science and Engineering\ (\bibinfo  {publisher} {Elsevier
  Science},\ \bibinfo {year} {1969})\BibitemShut {NoStop}%
\bibitem [{Note3()}]{Note3}%
  \BibitemOpen
  \bibinfo {note} {We have also assumed that the $\Delta _{i}$ do not differ by
  integer powers to avoid the complication of logarithmic terms in the
  Frobenius series represented by \protect \textup {\hbox {\mathsurround \z@
  \protect \normalfont (\ignorespaces \ref
  {Eq:GenericNearOriginExpansion}\unskip \@@italiccorr )}}.}\BibitemShut
  {Stop}%
\bibitem [{Note4()}]{Note4}%
  \BibitemOpen
  \bibinfo {note} {In fact, for $\lambda _{N}$ small enough the $\Delta _{i}$
  are real and one can find isolated bound states for some choices of the
  parameter $U(N)$.}\BibitemShut {Stop}%
\bibitem [{Note5()}]{Note5}%
  \BibitemOpen
  \bibinfo {note} {The Hamiltonian is to be made self-adjoint in the portion of
  the half-line considered.}\BibitemShut {Stop}%
\bibitem [{\citenamefont {Gitman}\ \emph
  {et~al.}(2012{\natexlab{b}})\citenamefont {Gitman}, \citenamefont {Tyutin},\
  and\ \citenamefont {Voronov}}]{9780817646622}%
  \BibitemOpen
  \bibfield  {author} {\bibinfo {author} {\bibfnamefont {D.}~\bibnamefont
  {Gitman}}, \bibinfo {author} {\bibfnamefont {I.}~\bibnamefont {Tyutin}}, \
  and\ \bibinfo {author} {\bibfnamefont {B.}~\bibnamefont {Voronov}},\ }\href
  {https://www.amazon.com/Self-adjoint-Extensions-Quantum-Mechanics-Applications-ebook/dp/B00F5QEXPM%3FSubscriptionId%3D0JYN1NVW651KCA56C102%26tag%3Dtechkie-20%26linkCode%3Dxm2%26camp%3D2025%26creative%3D165953%26creativeASIN%3DB00F5QEXPM}
  {\emph {\bibinfo {title} {Self-adjoint Extensions in Quantum Mechanics:
  General Theory and Applications to Schr\"{o}dinger and Dirac Equations with
  Singular Potentials: 62 (Progress in Mathematical Physics)}}}\ (\bibinfo
  {publisher} {Birkhäuser},\ \bibinfo {year} {2012})\BibitemShut {NoStop}%
\bibitem [{\citenamefont {Asorey}\ \emph {et~al.}(2006)\citenamefont {Asorey},
  \citenamefont {Garcia-Alvarez},\ and\ \citenamefont
  {Munoz-Castaneda}}]{Asorey:2006pr}%
  \BibitemOpen
  \bibfield  {author} {\bibinfo {author} {\bibfnamefont {M.}~\bibnamefont
  {Asorey}}, \bibinfo {author} {\bibfnamefont {D.}~\bibnamefont
  {Garcia-Alvarez}}, \ and\ \bibinfo {author} {\bibfnamefont {J.~M.}\
  \bibnamefont {Munoz-Castaneda}},\ }\bibfield  {booktitle} {\emph {\bibinfo
  {booktitle} {{7th Workshop on Quantum Field Theory Under the Influence of
  External Conditions (QFEXT 05) Barcelona, Catalonia, Spain, September 5-9,
  2005}}},\ }\href {\doibase 10.1088/0305-4470/39/21/S03} {\bibfield  {journal}
  {\bibinfo  {journal} {J. Phys.}\ }\textbf {\bibinfo {volume} {A39}},\
  \bibinfo {pages} {6127} (\bibinfo {year} {2006})},\ \Eprint
  {http://arxiv.org/abs/hep-th/0604089} {arXiv:hep-th/0604089 [hep-th]}
  \BibitemShut {NoStop}%
%%CITATION = HEP-TH/0604089;%%
\end{thebibliography}%

% %%%%%%%%%%%% Supplementary Material %%%%%%%%%%%%
% 
\clearpage


\section*{Supplementary material}

# Scale anomaly of a Lifshitz scalar: a universal quantum phase transition to discrete scale invariance
## Supplementary materials

Daniel K. Brattan[*]
*Interdisciplinary Center for Theoretical Study, University of Science and Technology of China, 96 Jinzhai Road, Hefei, Anhui, 230026 PRC.*

Omrie Ovdat[†] and Eric Akkermans[‡]
*Department of Physics, Technion Israel Institute of Technology, Haifa 3200003, Israel.*
(Dated: April 21, 2018)

## SELF-ADJOINT EXTENSION

### The eigenfunctions

The energy eigenvalue equation with Hamiltonian

$$\hat{H}_N = (-1)^N \frac{d^{2N}}{dx^{2N}} - \frac{\lambda_N}{x^{2N}} \ . \tag{S1}$$

has an analytic solution in terms of generalised hypergeometric functions [S1] for arbitrary energy and $\lambda_N$. These analytic solutions are specified by the roots of the indicial equation,

$$\lambda_N = (-1)^N \prod_{k=0}^{2N-1} (\Delta - k) \ , \tag{S2}$$

which can be obtained by inserting $\Psi(x) = x^\Delta (1 + \mathcal{O}(x^{2N}))$ into the eigenvalue equation and solving for $\Delta$. We label the $2N$ solutions of (S2) as $\Delta_i, i = 1, \ldots, 2N$ and order by ascending real part followed by ascending imaginary part. These roots can be collected into pairs which sum to $2N-1$. The $\Delta_i$ also allow us to rewrite the differential operator as

$$x^{2N} \hat{H}_N = (-1)^N \prod_{j=1}^{2N} \left( x \frac{d}{dx} - \Delta_{\sigma(j)} \right) \ , \tag{S3}$$

where $\sigma$ is an arbitrary permutation of the integers $1, \ldots, 2N$.

Let $\boldsymbol{\Delta}_i$ be the vector of solutions to (S2) with $\Delta_i$ omitted, then generic solutions of the differential equation associated with $\hat{H}_N$ and energy eigenvalue $E$ are given by the sum:

$$\Psi(x; \phi_j) = \sum_{i=1}^{2N} e^{i\Delta_i \phi_j} \left( \frac{\epsilon x}{2N} \right)^{\Delta_i} \Gamma \left( \frac{\boldsymbol{\Delta}_i - \Delta_i}{2N} \right)$$
$${}_0F_{2N-1} \left[ \begin{array}{c} - \\ 1 - \frac{\boldsymbol{\Delta}_i - \Delta_i}{2N} \end{array} ; E \left( \frac{x}{2N} \right)^{2N} \right] \ , \tag{S4}$$

$$\phi_j = \frac{1}{2N} (\theta - 2\pi(N + (w + j))) \ ,$$
$$j = 1, 2, \ldots, 2N \ , \tag{S5}$$

where $w$ is an integer chosen such that $|\phi_j| < \pi(1 + \frac{1}{2N})$, $E = |E|e^{i\theta}$ and $\epsilon = |E|^{\frac{1}{2N}}$. For negative energies the choices of $\phi_j$ are symmetric about zero, which may or may not be included in the set. The functions ${}_0F_{2N-1}$ are generalised hypergeometric functions [S1].

We now wish to determine the wavefunction corresponding to a $U(1)$ self-adjoint extension at sufficiently positive $\lambda_N$, i.e. where there are only $N+1$ normalisable roots. To do this, consider a general decaying wavefunction which is given by a sum over the $N$ exponentially decaying basis wavefunctions defined in (S4):

$$\Psi(x) = \sum_{j=1}^{N} A_j \Psi(x; \phi_j) \ , \tag{S6}$$

where $A_j$ are some arbitrary complex constants defining the wavefunction. Without loss of generality let $\Delta_1$ have a real part less than $-1/2$. To remove this root from our general expression we require that

$$\sum_{j=1}^{N} A_j e^{i\Delta_1 \phi_j} \left( \frac{\epsilon x}{2N} \right)^{\Delta_1} \Gamma \left( \frac{\boldsymbol{\Delta}_1 - \Delta_1}{2N} \right)$$
$${}_0F_{2N-1} \left[ \begin{array}{c} - \\ 1 - \frac{\boldsymbol{\Delta}_1 - \Delta_1}{2N} \end{array} ; E \left( \frac{x}{2N} \right)^{2N} \right] \equiv 0 \ , \tag{S7}$$

near $x = 0$. We can drop all the $j$ independent terms leaving

$$\sum_{j=1}^{N} A_j e^{i\Delta_1 \phi_j} = 0 \ . \tag{S8}$$

The $A_j$ are constrained. Similarly, for the other $N-2$ non-normalisable roots in the $U(1)$ regime which we can summarise by the matrix equation:

$$\left( e^{i\Delta_i \phi_j} \right) A_j = 0 \ . \tag{S9}$$

We note that the matrix is of dimension $(N-1) \times N$ and, as the rows are generically linearly independent, the vector $A_j$ parameterizes the one-dimensional null space.

Let $a_j$ denote the components of this unit normalised vector that solves (S9). The desired $U(1)$ wavefunction

is then given by

$$\Psi(x) = A \sum_{i=1}^{N} \left(\frac{\epsilon x}{2N}\right)^{\Delta_i} \left(\sum_{j=1}^{N} a_j e^{i\Delta_i \phi_j}\right) \Gamma\left(\frac{\mathbf{\Delta}_i - \Delta_i}{2N}\right)$$

$$_0F_{2N-1}\left[\begin{array}{c} - \\ 1 - \frac{\Delta_i - \mathbf{\Delta}_i}{N} \end{array}; E\left(\frac{x}{2N}\right)^{2N}\right] \quad (S10)$$

where $A$ is an overall normalisation constant and the energy is completely arbitrary. To fix the energy we require a boundary condition from $x = 0$ as discussed in the main text. The details follow as discussed there.

## CUT-OFF REGULARISATION

### Relation to zero cut-off

The full space of self-adjoint extensions in the cut-off problem is larger than or equal to the parameter space at zero cut-off. For a non-zero cut-off $x_0 \neq 0$ and for all $\lambda_N \in \mathbb{R}$, there is a $U(N)$ parameter space of boundary conditions at $x = x_0$. However, for the zero cut-off problem, the dimension of the space of self-adjoint extensions varies as a function of $\lambda_N$ as can be seen from table I. As a result, in the limit $x_0 \to 0$, the dimension of the parameter space of self-adjoint extensions with $x_0 \neq 0$ may not match onto that at zero cut-off. This occurs already for $N = 1$. For $\lambda_1 > -3/4$, there is a unique self-adjoint extension when there is no cut-off. However, introduction of a cut-off allows an infinite $U(1)$ set of boundary conditions. The difference between the zero cut-off and non-zero cut-off cases is due to the fact that, for all $x_0 \neq 0$, there is no issue of divergence and normalisability as $x \to 0$ because this region is excluded. Therefore, unlike the zero cut-off case, the self-adjoint extension space of $x_0 \neq 0$ is bigger than or equal to that of $x_0 = 0$ since it allows the existence of solutions that are not normalisable as $x \to 0$.

### Constraints on the cut-off boundary conditions

We can diagonalize the boundary form on the cut-off by defining $\Psi_k^{\pm}(x_0)$,

$$x_0^{k-1} d_x^{k-1} \Psi(x_0) = \Psi_k^+(x_0) + \Psi_k^-(x_0) , \quad (S11)$$
$$x_0^{2N-k} d_x^{2N-k} \Psi(x_0) = e^{i\pi(k-\frac{1}{2})} \left[\Psi_k^+(x_0) - \Psi_k^-(x_0)\right] \quad (S12)$$

for $1 \leq k \leq N$. After this redefinition we find the boundary form is given by:

$$\frac{1}{2}(ix_0)^{2N-1} [\Phi, \Psi](x_0)$$
$$= \mathbf{\Phi}^+(x_0)^\dagger \cdot \mathbf{\Psi}^+(x_0) - \mathbf{\Phi}^-(x_0)^\dagger \cdot \mathbf{\Psi}^-(x_0) . \quad (S13)$$

The vanishing of the boundary form [S2, S3] can be achieved by setting

$$\mathbf{\Psi}^+(x_0) = U_N \mathbf{\Psi}^-(x_0) , \quad \mathbf{\Phi}^+(x_0) = U_N \mathbf{\Phi}^-(x_0) . \quad (S14)$$

It is well-known that a generic self-adjoint extension does not preserve time-reversal invariance [S4], nor does it guarantee that the spectrum is bounded from below. The former property is equivalent to having real-valued ground state wavefunctions [S4] and can be imposed on the system by limiting the self-adjoint extensions to those which satisfy

$$\det\left(\mathbb{1}_N - U_N^* U_N\right) = 0 . \quad (S15)$$

To bound the spectrum from below consider the expectation value of the energy on some unit normalised state $|\Psi\rangle$:

$$\langle\Psi|\hat{H}_N|\Psi\rangle = \sum_{j=1}^{N} (-1)^{N+j-1} d_{x_0}^{j-1} \Psi^*(x_0) d_{x_0}^{2N-j} \Psi(x_0)$$
$$+ \int_{x_0}^{\infty} dx \left[|\partial_x^N \Psi|^2 - \frac{\lambda_N}{x^{2N}}|\Psi|^2\right] .$$

Using the redefinitions of (S11) one can show that $\langle\Psi|\hat{H}_N|\Psi\rangle$ is greater than or equal to

$$\frac{1}{x_0^{2N}} \left[(-1)^N x_0 \left(\mathbf{\Psi}^-(x_0)\right)^\dagger \left[i\left(U_N - U_N^\dagger\right)\right] \mathbf{\Psi}^-(x_0)\right.$$
$$\left. - \lambda_N\right] . \quad (S16)$$

Thus, if the Hermitian part of the unitary matrix has eigenvalues of the correct sign we can ensure that the spectrum is bounded from below by the minimum of the potential [S5].

### Example: cut-off regularisation energies for $N = 1$

In the case of $N = 1$ it is known that for $\lambda_1 > \frac{1}{4}$, if the logarithmic derivative of the wavefunction is chosen to be any real number one finds at low energies discrete scale invariance [S6]. This classic result follows by noting that (S13) implies

$$\frac{x_0 \Psi'(x_0)}{\Psi(x_0)} = \tan\left(\frac{\theta}{2}\right) \quad \theta \neq \pm\pi , \quad (S17)$$

as a rewriting of the boundary condition. The cases of $\pm\pi$ correspond to Dirichlet and Neumann conditions for the wavefunction at the cut-off and can be thought of as limits. Taking small energies is equivalent to taking the cut-off to zero where for $N = 1$ the wavefunction has the

| $N=1$ | $-3/4 < \lambda_1$ | $\lambda_1 < -3/4$ | | |
|---|---|---|---|---|
| extension parameter | $U(1)$ | essentially self-adjoint | | |
| $N=2$ | $105/16 < \lambda_2$ | $-45 < \lambda_2 < 105/16$ | $\lambda_2 < -45$ | |
| extension parameter | $U(1)$ | $U(2)$ | essentially self-adjoint | |
| $N=3$ | $693 \lesssim \lambda_3$ | $-162 \lesssim \lambda_3 \lesssim 693$ | $-36201 \lesssim \lambda_3 \lesssim -162$ | $\lambda_3 \lesssim -36201$ |
| extension parameter | $U(1)$ | $U(3)$ | $U(2)$ | essentially self-adjoint |

TABLE I. A table showing the distinct regimes of self-adjoint extension parameter for $N = 1, 2, 3$. The bounds for $N = 1, 2$ are exact while those for $N = 3$ are approximate and determined by numerically solving the indicial equation (S2).

form:

$$\Psi(x_0) = \tilde{A} \left(\frac{\epsilon x_0}{2}\right)^{\frac{1}{2}} |\Gamma(-i\nu_1)|^{\frac{1}{2}}$$
$$\cos\left(\nu_1 \ln\left(\frac{\epsilon x_0}{2}\right) + \frac{\phi_1}{2}\right) + \mathcal{O}^{\frac{3}{2}}(x_0) \,, \quad (S18)$$

$$e^{i\phi_1} = \frac{\Gamma(-i\nu_1)}{\Gamma(i\nu_1)} \,, \qquad \nu_1 = \sqrt{\lambda_1 - \frac{1}{4}} \,, \quad (S19)$$

with $\tilde{A}$ some normalisation constant. Substituting (S18) into (S17) we find

$$\tan\left(\frac{\theta}{2}\right) = \sqrt{\lambda_1 - \frac{1}{2}} \frac{\cos\left(\nu_1 \ln\left(\frac{\epsilon x_0}{2}\right) + \frac{\phi_1 + \alpha}{2}\right)}{\cos\left(\nu_1 \ln\left(\frac{\epsilon x_0}{2}\right) + \frac{\phi_1}{2}\right)} \,, \quad (S20)$$

$$e^{i\alpha} = \frac{\frac{1}{2} + i\nu_1}{\frac{1}{2} - i\nu_1} \,. \quad (S21)$$

Fixing $\theta$ we can numerically solve this equation for an $\epsilon$. Taking this as a reference energy, the symmetry of the right hand side ensures that $\epsilon \exp(-\pi n/\nu_1)$ is also a solution where $n \in \mathbb{N}$ is required for the approximation (S18) to apply. Hence, for $N = 1$ the cut-off regularisation gives approximate discrete scale invariance at low energies.

### Relating the energies to the critical coupling

By examining (S2) it can be seen that for $\lambda_N > \lambda_{N,c} \equiv (2N-1)!!^2/2^{2N}$ there is a pair of complex roots whose real part is fixed to be $N - 1/2$ i.e.

$$\Delta_N = \Delta_{N+1}^* = (N - 1/2) - i\nu_N(\lambda_N) \,, \quad (S22)$$
$$\nu_N(\lambda_N) = \sqrt{\lambda_N - \lambda_{N,c}} \left(\alpha_N^{-1} + \mathcal{O}(\lambda_N - \lambda_{N,c})\right) \,, \quad (S23)$$

where $\alpha_N$ is related to the value of the second derivative of the indicial equation (S2) with respect to $\Delta$ at $\Delta = N - 1/2$. The precise form of $\alpha(N)$ is given by:

$$\alpha_N^2 = \frac{\lambda_{N,c}}{2} \left(\gamma_1\left(N + \frac{1}{2}\right) - \gamma_1\left(\frac{1}{2} - N\right)\right) \,. \quad (S24)$$

---

* danny.brattan@gmail.com
† somrie@campus.technion.ac.il
‡ eric@physics.technion.ac.il

[S1] Y. Luke, *The Special Functions and Their Approximations*, Mathematics in Science and Engineering (Elsevier Science, 1969).
[S2] D. M. Gitman, I. V. Tyutin, and B. L. Voronov, (2009), 10.1088/1751-8113/43/14/145205, arXiv:0903.5277 [quant-ph].
[S3] D. Gitman, I. Tyutin, and B. Voronov, *Self-adjoint Extensions in Quantum Mechanics: General Theory and Applications to Schrödinger and Dirac Equations with Singular Potentials: 62 (Progress in Mathematical Physics)* (Birkhuser, 2012).
[S4] G. Bonneau, J. Faraut, and G. Valent, Am. J. Phys. **69**, 322 (2001), arXiv:quant-ph/0103153 [quant-ph].
[S5] We should emphasise that the implication is one-way i.e. the energies can still be bounded below if the Hermitian part of the unitary matrix does not eigenvalues of the correct sign but this must be determined by examining the energy spectrum.
[S6] A. M. Essin and D. J. Griffiths, American Journal of Physics **74**, 109 (2006).



\end{document}